\documentclass[journal]{IEEEtran}     % twocolumn
\usepackage{amsthm}

\usepackage{graphics,graphicx,epsfig}
\usepackage{amsfonts}
\usepackage{amsmath}
\usepackage{amssymb}
\usepackage{times}
\usepackage{setspace}
\usepackage[english]{babel}
\usepackage{pifont,array}
\usepackage[latin1]{inputenc}
\newtheorem{theorem}{Theorem}

\newtheorem{proposition}[theorem]{Proposition}

\newcommand{\eps} {\varepsilon}
\newcommand{\beps}{\boldsymbol{\varepsilon}}
\newcommand{\bbeta}{\boldsymbol{\beta}}

\newcommand{\cam}{\mathcal{M}}

\newcommand{\cah}{\mathcal{H}}
\newcommand{\ind}{1\!\!1}
\newcommand{\fa}{\mathsf{fa}}
\newcommand{\ca}{\mathsf{ca}}

\begin{document}

\title{On the Effect of Data Contamination on Track Purity}

\author{\IEEEauthorblockN{Adrien Ickowicz\IEEEauthorrefmark{1}, and J.-Pierre Le Cadre\IEEEauthorrefmark{2}}\\
~\\
\IEEEauthorblockA{\IEEEauthorrefmark{1} CSIRO Mathematics, Informatics, Statistics, Locked Bag 17, North Ryde 1670 NSW, Australia}\\
\IEEEauthorblockA{\IEEEauthorrefmark{2} IRISA/CNRS, Campus de Beaulieu, 35042, Rennes~ (cedex), France}\\
Corresponding author: A. Ickowicz (e-mail: adrien.ickowicz@csiro.au)}
%\author{
%Adrien Ickowicz and J.-Pierre Le Cadre,\\
%IRISA/CNRS\\
%Campus de Beaulieu,\\
%35042, Rennes~ (cedex), France \thanks{e-mail: ickowicz,lecadre@irisa.fr}\\}

\maketitle

\thispagestyle{empty}

\begin{abstract}
This paper is concerned with performance analysis for data association, in a target tracking environment. Effects of misassociation are considered in a simple (linear) multiscan framework so as to provide closed-form expressions of the probability of correct association. In this paper, we focus on the development of explicit approximations of this probability. Via rigorous calculations the effect of dimensioning parameters (number of scans, false measurement positions or densities)  is analyzed, for various modelings of the false measurements. Remarkably, it is possible to derive very simple expressions of the probability of correct association which are independent of the scenario kinematic parameters.
\end{abstract}
\vspace{-.5 cm}
{\small
\section*{Index of principal notations:}
\begin{itemize}
\item{${\sf ca}$: correct association~,~${\sf fa}$: false association, DTMC: discrete time Markov chain.}
\item{ ${\sf erfc}(x)=\int_{x}^{+\infty} {\mathcal{N}}(0,1)(x)\,dx$~, ~${\mathcal{N}}(m,\sigma)$: normal density mean $m$, s.d. $\sigma$.}
\item{$I$: identity matrix~,~${\mathbf 1}$: indicator function~,~$\ind$: a vector made of $1$.}
\item{$N$: scan number, $l$: a scan index, $\lambda$: the false alarm distance.}
\item{$\Delta_{f,c}$: difference of association costs~,~$K$: number of false measurements.}
\end{itemize} }
\vspace{-.5 cm}
\section{Introduction}
A fundamental problem in multi-target tracking is to evaluate the performance of the association algorithms. However, it is quite obvious that tracking and association are completely entangled. In this context, a key performance measure is the probability of correct association. Generally, track accuracy has been considered without consideration of the association problem. However, remarkable exceptions exist. Very roughly, they can be divided in two categories. The first one deals with track divergence. In particular, important efforts have been done for performance of the Nearest Neighbor (NN) filter. In some approaches, the tracking error is modeled as a diffusion process \cite{rog}. Fundamental contributions deals with the analysis of the dynamic process of tracking divergence \cite{ber}, applied to NN filter performance\cite{lib1} or the expected track life of the PDAF \cite{bar-tse} in clutter \cite{lib2}. Equally important are contributions devoted to the performance evaluation of track initiation in dense environments \cite{bar-li}, \cite{chan}. \\

The second category is scan-wise oriented, which means that for each set of measurements, the algorithm calculates an optimal track-to-measurement assignment and propagate only the best "hypothesis". Since it uses an optimal track-to-measurement assignment it should provide better tracking performance than NN or PDA \cite{mor1}, \cite{mor2}. However, this work is essentially oriented toward a modeling of misassociations via the effect of permutations, from a $0$-scan viewpoint and its propagation \cite{chan}. Here, we focus on the effect of the "contamination" of a target track due to extraneous measurements, within a multiscan framework. In fact, a "contamination" results in a change of the estimates of the track parameters, which could render misassociations more likely than the true one. It is certain that only measurements situated in the immediate vicinity of the target track would have a severe effect. This is the case for dense target environment or for situations where these close outliers are intentionally generated (e.g. decoys) \cite{slw}.\\

Here, our analysis is devoted to multiscan association analysis. For easing calculations the target motion is generally assumed to be deterministic, while we are concerned with batch performance.  The linear estimation framework has been used so as to allow us to obtain explicit closed-form expressions of the probability of correct association, which is the only aim of this contribution. Then, track purity can be seen as the probability that the proportion of false measurements "included" in the system track be under a certain level (percentage). False measurements are modeled either as deterministic or random. \\

 This paper is organized as follows. In Section $2$ the elementary multiscan association scenario is presented. We have then to calculate the  association costs under the two hypotheses (correct and false associations). This is the object of Section 3. The major result of this section is the calculation of (exact) closed-forms for these association costs via elementary linear algebra, which will be of constant use subsequently.\\

The true problem is now to derive from Section 3 results an accurate closed-form approximation of the probability of correct association. This is precisely the aim of Section 4, which plays the central role in this paper. The way we derive this approximation is detailed. It is based upon an approximation of the normal density via a sum of indicator (step) functions. The final result is a very simple closed-form approximation, whose accuracy is testified by Section 5 (simulation results). Note, however, that these results are limited to a single false association within the whole batch period. \\

It is the aim of Section 6 to extend the analysis to multiple false measurements. The approach we developed for approximating the probability of correct association in the unique false measurement case  is no longer valid. In particular,the method we used for approximating the integrals no longer holds. So, we have to resort to a different approach. Roughly, we consider that the mean and variance of the difference of association costs are characterized by their distributions, themselves depending on random parameters. It is shown that the probability of correct association is highly dependent of the number of false measurements lying in the vicinity of the target trajectory.

\section{Problem formulation}

A target is moving with a rectilinear and uniform motion. Noisy measurements consisting of Cartesian positions are represented by the points:

\begin{equation}
 \tilde{P}_{1}=\left(\tilde{x}_{1},\tilde{y}_{1}\right)\;,\;\tilde{P}_{2}=\left(\tilde{x}_{2},\tilde{y}_{2}\right),\cdots,\;\tilde{P}_{N}=\left(\tilde{x}_{N},\tilde{y}_{N}\right) \;,
\end{equation} 
 at time periods $t_{1}$, $t_{2}$,$\cdots$,$t_{N}$, which are called  "{\it scans}". Under the correct association hypothesis, the position measurements are the exact Cartesian positions $P_{i}=(x_{i},y_{i})$, corrupted by a sequence of independent and identically normally distributed noises (denoted $\eps_{ x_{i}},\;\eps_{y_{i}}$), i.e.:

\begin{equation}
\tilde{P}_{i}=\left(\tilde{x}_{i},\,\tilde{y}_{i}\right)=\left(x_{i}+\eps_{x_{i}},\,y_{i}+\eps _{y_{i}}\right)\;.
\end{equation} 
We assume that the observation noises $\eps_{x}$ and $\eps_{y}$ are uncorrelated, with a variance $\sigma^{2}$. When a target is ({\it sufficiently}) isolated from others, there is no ambiguity about the measurement origin. This is not true if a second target lies in the vicinity of the first target. In this case, it becomes possible to make a mistake about the origin of an observation by associating it to the wrong target, thus corrupting target trajectory estimation. But the question is to give a more precise meaning to the term "sufficiently isolated".\\

\begin{figure}[ht!]
\centering
\scalebox{0.40}{\includegraphics{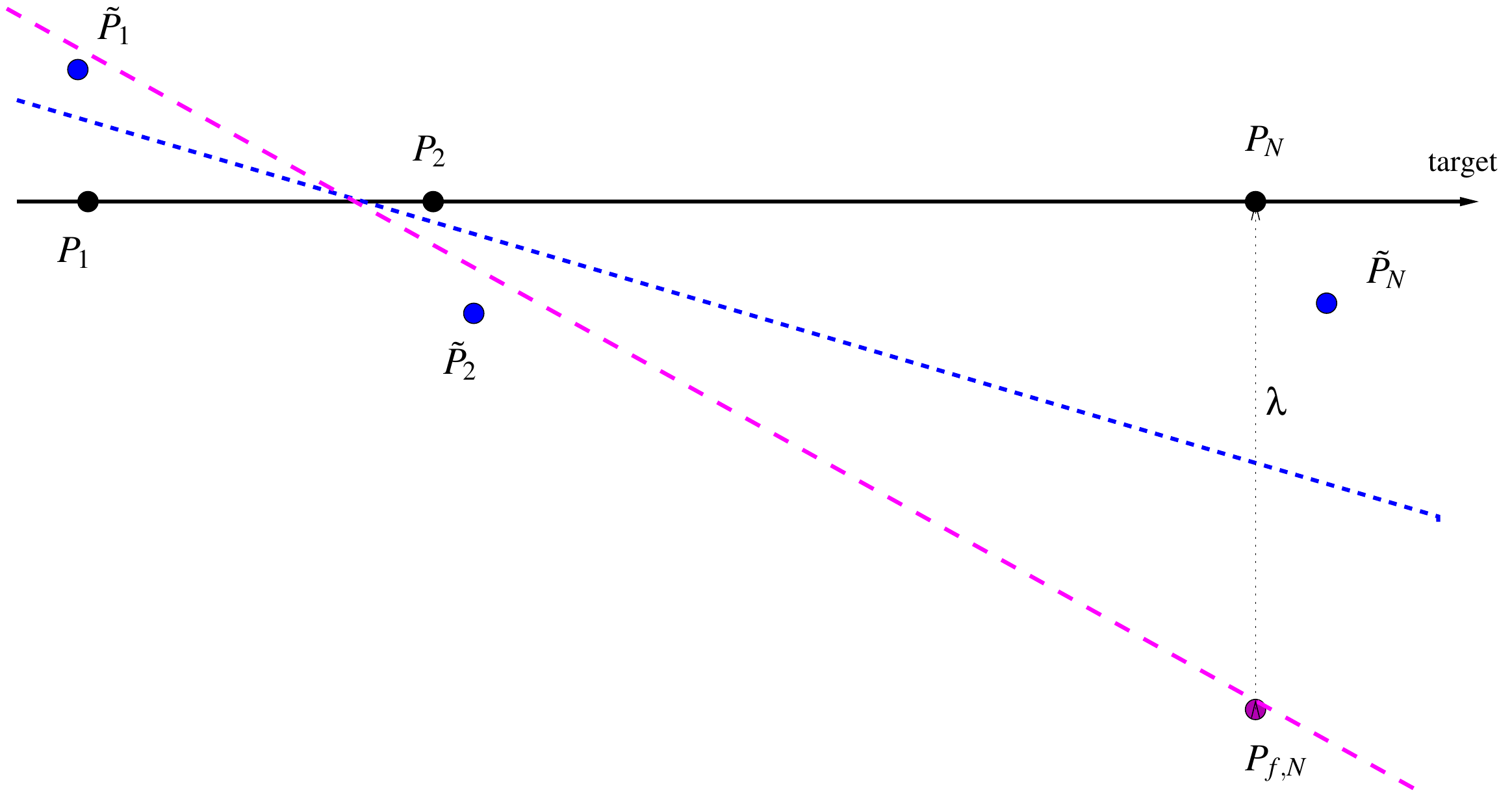}}
\caption{\it The  association scenario. Dotted line: correct association, dashed line: false association. }
\label{scenar}
\end{figure}

Thus, the aim of this article is to give a closed-form expression for the probability of correct association of  measurements to a target track, as a function of the number of scans and the distance of the outliers observations. In order to simplify the scenario, we consider that the outlier measurements $P_{f}$ are located close to the true  target position $P_{l}=(x_{l},y_{l})$ at time period $t_{l}$, with a distance $\lambda$\footnote{For the sake of brevity, we assume that measurements are resolved (see \cite{chanb})}. Throughout this paper $\lambda$ stands for the ratio $\lambda/\sigma$. The general problem setting and definitions are depicted in fig.~\ref{scenar}. \\
Let us denote $\delta=t_{i+1}-t_{i}$, the inter-measurement time, and:
$$
{\bf{v}}=\left(v_{x},v_{y}\right)^{T}\;,
$$
the two components of the constant target velocity on the Cartesian axis. Then, in the deterministic case, the target trajectory is  defined by the  state vector $\left(x_{1},y_{1},v_{x},v_{y}\right)$. 

\section{Problem analysis }

Under the correct association ({\sf ca}) hypothesis and denoting $\tau_{i}\stackrel{\Delta}{=}i\;\delta$, the  position measurements $\tilde{P}_{i}$ are represented by the following equation\footnote{$I$: identity matrix}:

\begin{equation}
\underbrace{
\left(
\begin{array}{c}
\tilde{x}_{1}\\
\tilde{y}_{1}\\
\tilde{x}_{2}\\
\tilde{y}_{2}\\
\vdots\\
\tilde{x}_{N}\\
\tilde{y}_{N}
\end{array}
\right)}_{\tilde{Z}_{\sf{ca}}}
 =
 \underbrace{
\left(
\begin{array}{cc}
I_{2} & 0_{2}\\
I_{2} & \tau_{1}I_{2}\\
\vdots & \vdots\\
I_{2} & \tau_{N-1}I_{2}
\end{array}
\right)}_{\mathcal{X}}
\;
\underbrace{
\left(
\begin{array}{c}
x_{1}\\
y_{1}\\
v_{x}\\
v_{y}
\end{array}
\right) }_{\bbeta}
+
\underbrace{
\left(
\begin{array}{c}
\varepsilon_{x_{1}}\\
\varepsilon_{y_{1}}\\
\varepsilon_{x_{2}}\\
\varepsilon_{y_{2}}\\
\vdots\\
\varepsilon_{x_{N}}\\
\varepsilon_{y_{N}}
\end{array}
\right)}_{\tilde{\beps}_{\sf{ca}}}
\label{determ_1}
\end{equation}
 With these definitions and under the correct association hypothesis, the measurement model simply stands as follows:
\begin{equation}
\tilde{Z}_{{\sf ca} }  = {\mathcal X} \; \bbeta + \tilde{\beps}_{{\sf ca}} \;.
\end{equation} 

\subsection{The regression model \cite{bav}}

Consider the following linear regression model:

\begin{equation}
\tilde{Z}={\mathcal X}\; \bbeta + \tilde{\beps}\;,
\end{equation} 
where $\tilde{Z}$ are the data, ${\mathcal X}$ are the regressors and $\bbeta$ is the vector of parameters, to be estimated. Generally, the estimation of $\bbeta$ is made via the quadratic loss function:

\begin{equation}
{\mathcal L}_{2}(\bbeta)=\left(\tilde{Z}-{\mathcal X}\;\bbeta\right)^{T}\,\left(\tilde{Z}-{\mathcal X}\;\bbeta\right)=\Vert \tilde{Z}-{\mathcal X}\;\bbeta \Vert ^{2} \;.
\end{equation} 
If the matrix ${\mathcal X}^{T}{\mathcal X}$ is non-singular, then ${\mathcal L}_{2}(\bbeta)$ is minimum for the unique value $\hat{\bbeta}$ of $\bbeta$ such that:

\begin{equation}
\hat{\bbeta}=({\mathcal X}^{T}\;{\mathcal X})^{-1}{\mathcal X}^{T}\,\tilde{Z} \;.
\end{equation} 
From the estimation $\hat{\bbeta}$ of $\bbeta$, let  $\widehat{Z}$ be the estimator of the mean ${\mathcal X}\; \bbeta$ of the random vector $\tilde{Z}$ defined by:
\begin{equation}
\begin{array}{lll}
\widehat{Z} &= & {\mathcal H}\;\tilde{Z}\;,\\ \nonumber
\mbox{with :}& & \\
{\mathcal H}& = & {\mathcal X}({\mathcal X}^{T}{\mathcal X})^{-1}\;{\mathcal X}^{T} \;.
\end{array}
\label{reg-H}
\end{equation} 
 The vector of the residuals $\hat{\beps}\stackrel{\Delta}{=}\tilde{Z}-\widehat{Z}$ is given by:
\begin{equation}
\hat{\beps} =\mathcal{M}\;\tilde{Z}\;,
\end{equation} 
with $\mathcal{M}= I -\mathcal{H}$ , and $I$  the identity matrix. It is easy to check that $\mathcal{M}$ is a projection matrix (i.e. $\mathcal{M}^{T}=\mathcal{M} $ and ${\mathcal{M}}^{2}=\mathcal{M}$). We also recall the following classical identities, which will be used subsequently \cite{ant}:

\begin{equation}
\mathcal{M}\;\mathcal{X}  =  0\;,\;\;\mbox{and:} \;
\hat{\beps} = \mathcal{M}\; \tilde{\beps} \;.
\label{epsilon_chap}
\end{equation} 

\subsection{Evaluation of the correct association probability}

Assume that the outlier measurement $P_{f,l}=(x_{f},y_{f})$ is located at the point ($1\leq l\leq N$, see fig.~\ref{scenar}):

\begin{equation}
\left\{
\begin{array}{l l l}
x_{f} & = & x_{l} \;, \\ \nonumber
y_{f} & = & y_{l}-\lambda \;.
\end{array}
\right.
\end{equation}
\noindent The correct association is then defined by the association of points $\left\{\tilde{P}_{1},\cdots,\tilde{\mathbf{P}}_{l},\cdots,\tilde{P}_{N}\right\} \stackrel{\Delta}{=}\widetilde{Z}_{\sf ca} $, whereas the wrong association is defined by $\left\{\tilde{P}_{1},\cdots,\tilde{\mathbf{P}}_{f,l},\cdots,\tilde{P}_{N} \right\}\stackrel{\Delta}{=}\widetilde{Z}_{\sf fa} $ (the lowercase $f$ stands for false association). The vectors $\widehat{Z}_{\sf ca}$ and $\widehat{Z}_{\sf fa}$ are similarly defined from $\tilde{Z}_{\sf ca}$, $\tilde{Z}_{\sf fa}$ and the regression equation (eq. \ref{reg-H}).\\

The vectors of residuals are $\hat{\beps}_{\sf ca}=\tilde{Z}_{\sf ca}-\widehat{Z}_{\sf ca}$ under the correct association hypothesis ($\sf ca$) and $\hat{\beps}_{\sf{fa}}=\tilde{Z}_{\sf fa}-\widehat{Z}_{\sf fa}$ under the false association hypothesis (${\sf{fa}}$). They are deduced from a linear regression, leading to the following definition of the costs of correct association (denoted $\mathcal{C}_{\sf ca}$)  and false association (denoted $\mathcal{C}_{\sf{fa}}$) :
\begin{eqnarray}
{\mathcal C}_{\sf ca} & = & (\tilde{Z}_{\sf ca}-\widehat{Z}_{\sf ca})^{T}\;(\tilde{Z}_{\sf ca}-\widehat{Z}_{\sf ca} ) \;,\\ \nonumber
 & = & \tilde{\beps}_{\sf ca}^{T} \;\mathcal{M} \;{\tilde{\beps}}_{\sf ca} \;.
\end{eqnarray} 
In the same way, we have also:

\begin{equation}
{\mathcal C}_{\sf{fa}}= \tilde{\beps}_{\sf{fa}}^{T} \;\mathcal{M}\; {\tilde{\beps}}_{\sf{fa}} \;.
\end{equation}
Let us define now $\Delta_{f,c}$ the difference between the correct and wrong costs, i.e.:

\begin{equation}
\label{dcf}
\Delta_{f,c}\stackrel{\Delta}{=} {\mathcal C}_{\sf{fa}}-{\mathcal C}_{\sf{ca}} \;.
\end{equation} 
Then, the probability of correct association is defined by the probability that $\Delta_{f,c} \geq 0$ (denoted $P(\Delta_{f,c}\geq 0)$). {\bf The aim of this article is to give closed-form expressions for this probability}.

Let be $\tilde{\beps}_{\sf{com}}$ the vector of components that the  vectors $\tilde{\beps}_{\sf{ca}}$ and $\tilde{\beps}_{\sf{fa}}$ have in common,
and define $\tilde{\beps}_{l}$ and ${\sf{fa}}_{l}$ as the complementary vectors \footnote{ This means that vectors  $\tilde{\beps}_{l}$ and ${\sf{fa}}_{l}$ are made of zeroes, excepted in the $l$ positions}, so that:

\begin{equation}
\tilde{\beps}_{\sf{ca}}=\tilde{\beps}_{\sf{com}}+\tilde{\beps}_{l} \;\;,\;\;\tilde{\beps}_{\sf{fa}}=\tilde{\beps}_{\sf{com}}+{\sf{fa}}_{l}\;.
\end{equation} 
With these notations, the difference between the correct and wrong costs $\Delta_{f,c}$  can be written:

\begin{equation}
\begin{array}{lll}
\medskip
\Delta_{f,c} & = & {\sf{fa}}_{l}^T \cam \;{\sf{fa}}_{l} - {(\tilde{\beps}_{l})}^T \cam (\tilde{\beps}_{l}) \;,\\
& & - 2\;{\left( \tilde{\beps}_{l}-{\sf{fa}}_{l} \right)}^T \cam {\left(\tilde{\beps}_{\sf{com}} \right)}\;.
\end{array}
\end{equation} 

 Since the components of the vector $\tilde{\beps}_{\sf{com}}$ are normally distributed and supposed independent, this vector is normal ($\tilde{\beps}_{\sf{com}} \sim  \mathcal{N}\left( \boldsymbol{O},{\Sigma}_{\sf{com}}\right)$ ), and similarly for $\tilde{\beps}_{l}$ ($\tilde{\beps}_{l} \sim {\mathcal N}\left(\boldsymbol{O},\Sigma_{l}\right)$ ).\\

Assuming that  the vector $\tilde{\beps}_{l}$ is set to a {\bf fixed} value $\mathbf{e}_{l}$, the law of the difference of costs $\mathcal{L}(\Delta_{f,c}\vert \,\tilde{\beps}_{l}= \mathbf{e}_{l})$ is normal with characteristics:
\begin{equation}
\begin{array}{l}
\mathcal{L}\left(\Delta_{f,c}\vert \,\tilde{\beps}_{l}= \mathbf{e}_{l} \right) = 
{\mathcal N}\left[ {\sf{fa}}_{l}^{T} \cam {\sf{fa}}_{l}-(\mathbf{e}_{l})^{T} \cam \mathbf{e}_{l},\, \right. \\
\left. \hspace{4cm} 4 (\mathbf{e}_{l}-{\sf{fa}}_{l})^{T} \Phi (\mathbf{e}_{l}-\sf{fa}_{l}) \right]\;,\\
\end{array}
\label{normal_1}
\end{equation} 
where: $\Phi \stackrel{\Delta}{=} \mathcal{M} \Sigma_{\sf{com}} \mathcal{M}^{T}$. Integrating this conditional density w.r.t. the Gaussian vector $\tilde{\eps}_{l}$, yields:
\begin{equation}
P(\Delta_{f,c}(l) \geq 0 )= \displaystyle{\mathbb{E}_{\tilde{\eps}_{l}}}\Bigg[{\sf {erfc}}\left(\frac{{\mathbf{e}_l}^T \cam \mathbf{e}_l-{\sf{fa}}_{l}^T \cam \;{\sf{fa}}_{l}}{2 \sqrt{(\mathbf{e}_l-{\sf{fa}}_{l})^T\Phi(\mathbf{e}_l-{\sf{fa}}_{l})}} \right)\Bigg]
\label{proba_expression_exacte}
\end{equation} 

Considering eq. \ref{proba_expression_exacte}, it is not surprising that it is the functional $\Psi(\mathbf{e}_l)$:

\begin{equation}
 \Psi(\mathbf{e}_l)= \frac{(\mathbf{e}_l)^T \cam \mathbf{e}_l-{\sf{fa}}_{l}^T \cam \;{\sf{fa}}_{l}}{2 \sqrt{(\mathbf{e}_l-{\sf{fa}}_{l})^T \Phi(\mathbf{e}_l-{\sf{fa}}_{l})}} \;,
\label{psi_1}
\end{equation} 
 which will play the fundamental role for analyzing the probability of correct association. However, though eq. \ref{proba_expression_exacte} is simple and general, it has the great inconvenient to involve the integration of the $\sf{erfc}$ function, so there is no hope to derive a closed-form expression of $P(\Delta_{f,c}(l) \geq 0 )$ by this way. So, we shall first turn toward a different approcah based on eq. \ref{normal_1}. To that aim, our developments follow the following steps:
\begin{itemize}
\item{Calculation of a closed form expression for the mean and variance of $\mathcal{L}\left(\Delta_{f,c}\vert \,\tilde{\beps}_{l}\right)$ (see eq. \ref{normal_1}) (see section \ref{matrix_1}).}
\item{Approximation of $\mathcal{L}\left(\Delta_{f,c}\vert \,\tilde{\beps}_{l}\right)$ as a sum of indicator functions, see section \ref{approx-indic}.}
\item{Approximation of the integration domains for the indicator functions, see section \ref{integ-domains}.}
\end{itemize}

\subsection{A closed-form for the  mean and variance of $\mathcal{L}\left(\Delta_{f,c}\vert \,\tilde{\beps}_{l} \right)$ }
\label{matrix_1}
Let us concentrate first on the case of a unique false association. Using elementary matrix calculations, the following results have been obtained (see Appendix A):

\begin{equation}
\begin{array}{lll}
\medskip
 {\sf{fa}}_{l}^{T} \cam {\sf{fa}}_{l}-(\mathbf{e}_{l})^{T} \cam \mathbf{e}_{l}&=& \left[\frac{2\left(2N+1-6l+\frac{6l^2}{N}\right)}{(N+1)(N+2)} -1\right]\;\left(\|\mathbf{e}_l\|^2 - \| {\sf{fa}}_{l} \|^2 \right)\;.\\ 
(\mathbf{e}_{l}-{\sf{fa}}_{l})^{T} \Phi (\mathbf{e}_{l}-\sf{fa}_{l})&=&\frac{1}{(N+1)^2(N+2)^2} 
\; \left[ Q_1(l,N) + 2l\;\delta Q_2(l,N) \right. \\
& & \left. \hspace{1cm} + l^2 \; \delta^2 Q_3(l,N) \right]\;{\|\mathbf{e}_l-{\sf{fa}}_{l}\|}^{2}\;.
\label{explicit_1}
\end{array}
\end{equation} 
where the $Q_1$, $Q_2$ and  $Q_3$ polynomials have the following expression:
\begin{equation}
\left|
\begin{array}{lll}
\medskip
Q_1(l,N) & =& 4N^3-50N^2+N(48l-18)+l(24-36l)+4 \;,\\  \nonumber
\medskip
Q_2(l,N) & = &-\frac{6}{\delta} \left[N^2-5N-2+4l(1+\frac{1}{N}-\frac{3l}{N})\right] \;  \\ \nonumber
\medskip
Q_3(l,N) & = & \frac{36}{\delta^2} \left[ \frac{N}{3}-1+\frac{2}{N}(\frac{1}{3}+2\,l-\frac{2\,l}{N^2}) \right] \;.
\end{array}
\right. 
\end{equation} 
Considering eq. \ref{proba_expression_exacte} (last row), we can notice that the variations of $\Psi(\mathbf{e}_l)$ as a function of $l$ are not very important. Actually, it is easily seen that $\frac{N^{2}} {2\;\left(N^{3}-3 l N^{2}+3 l^{2} N\right)^{1/2}} $ is varying between $\frac{\sqrt{N}}{2}$ and $ \frac{\sqrt{N}}{4}$ as $l$ varies between $0$ and $N$. Now, the ${\sf erfc}$ function is quite flat for large values of $N$, which means that $P(\Delta_{f,c}(\mathbf{e}_l) \geq 0)$ is almost  independent of the $l$ value.\\

The previous calculations can be rather easily extended to {\bf multiple} false associations.
Let ${\sf{FA}}_K = (l_k)_{k=1}^K$, be the vector made by indices $l_{k}$ of the (possible) false associations. A closed-form expression of the numerator of eq.~\ref{psi_1} is:
\begin{equation}
\begin{array}{l}
\hspace{-0.2cm} \mathbf{e}_{K}^T \cam \mathbf{e}_{K}-{\sf{FA}}_{K}^{T} \cam {{\sf{FA}}_{K}} =\displaystyle{\sum_{k=1}^K \sum_{k'=1}^K} \alpha_N(l_k,l_{k'}) \left( \langle\mathbf{e}_{l_k} , \mathbf{e}_{l_{k'}} \rangle - \langle\sf{fa}_{l_k} , \sf{fa}_{l_{k'}}\rangle \right)\;,\\ 
\mbox{with:}\\
\alpha_N(l_{k},l_{k'})=\left( {\mathbf{1}}_{\{k=k'\}}-\frac{2(2N+1-3\,l_{k'}-3\,l_k+\frac{6\,l_k l_{k'} }{N})}{(N+1)(N+2)} \right) \;.
\end{array}
\label{alp-thet}
\end{equation} 
Similarly, for the denominator $D_{\Psi_K}$ of $\Psi_{\sf{FA}_K}$, we have:
\begin{equation}
\begin{array}{l}
 {\left(\mathbf{e}_{K}- {\sf{FA}} \right)}^{T} \cam \left(\mathbf{e}_{K}- {\sf{FA}}\right)  = \\
\hspace{2cm} 2  \sqrt{\sum_{k=1}^{K}\sum_{k'=1}^{K} \theta(l_k,l_{k'}) \; \langle\mathbf{e}_{l_k}- {\sf{fa}}_{l_k}, \mathbf{e}_{l_{k'}}- {\sf{fa}}_{l_{k'} }}\rangle \;,\\
\mbox{with:} \\
(N+1)^2(N+2)^2\,\theta(l_k,l_{k'}) = 
\left[ Q_1^*({\sf{FA}}_K,N) \right. \\
\left. \hspace{2cm} +(l_k+l_{k'})\,Q_2^*({\sf{FA}}_{K},N) + {l_k}{l_{k'}} \,Q_3^*({\sf{FA}}_{K},N) \right]\;.
\end{array}
\label{dpsi}
\end{equation} 
The polynomials $Q_1^*$, $Q_2^*$ and $Q_3^*$ stand as follows:

\begin{equation}
\left|
\begin{array}{lll}
\medskip
Q_1^*({\sf{FA}}_{K},N) & = & \sum_{l=0, l\notin \sf{FA}_{K}}^N  (4N+2-6l)^2\;, \\\nonumber
\medskip
Q_2^*({\sf{FA}}_{K},N) & = & -\frac{6}{\delta} \;\left[\sum_{l=0, l\notin\sf{FA}_{K}} ^N    (4N+2-6l)(1-\frac{2 \,l}{N})\right] \;,   \\\nonumber
\medskip
Q_3^*({\sf{FA}}_{K},N) & = & \frac{36}{\delta^2} \;\left[\sum_{l=0,l\notin\sf{FA}_{K}} ^N  (1-\frac{2\,l}{N})^2 \right ] \;.
\end{array} 
\right.
\end{equation}

\section{Closed-form approximations of the probability of correct association: unique false measurement}

As shown in section $3$, it has been possible to obtain closed-form expressions of the $\Psi$ functional. However, even in the unique false measurement case, it is still necessary to perform an integration of the $\sf{erfc}(\Psi({\bf e}_{l})\;)$ functional. Though this is possible numerically, no analytic insight can be gained by this way. Actually, it is hopeless to consider approximations of the $\sf{erfc}$ function and we have to turn toward a radically different approach based on approximating the normal density by a sum of stepwise (indicator) functions.\\

\begin{figure}[ht!]
\centering
\scalebox{0.35}{\includegraphics{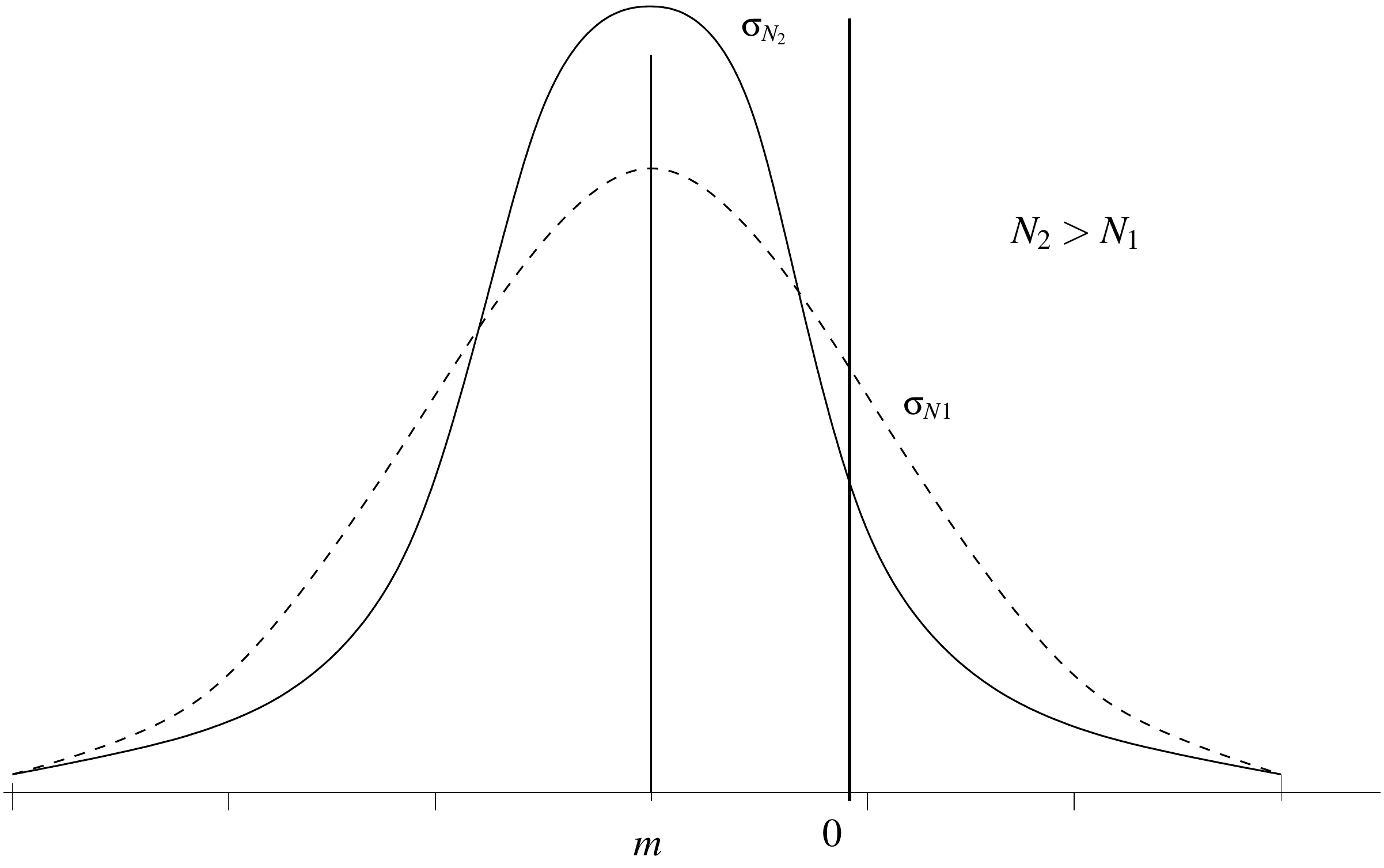}}
\caption{\it $P(\Delta_{f,c}\geq 0)$ as a function of $N$. }
\label{msigma}
\end{figure}

For the sake of simplicity, the error measurement components  $\tilde{\eps}_{x,l}$ and  $\tilde{\eps}_{y,l}$ will be simply denoted as $x$ and $y$. We have now to deal with convenient approximations of the association cost difference $\Delta_{f,c}\stackrel{\Delta}{=} {\mathcal C}_{\sf{fa}}-{\mathcal C}_{\sf{ca}}$. We restrict us to a single outlier measurement. At this point, it is worth recalling that it is {\it conditionally} distributed as a normal density (see eq. \ref{normal_1}):
\begin{equation}
\hspace{-0.5cm} \Delta_{f,c} \vert  \,\tilde{\beps}_{l}= \mathbf{e}_{l} \sim {\mathcal N}(m,\sigma).
\label{normalun}
\end{equation} 
The conditional mean $m$ and variance  $\sigma^{2}$ have been made explicit in section \ref{matrix_1} (eq. \ref{explicit_1}), yielding:
\begin{equation}
\left\{
\begin{array}{l}
\medskip
m=\left[\frac{2\left(2N+1-6 \,l+\frac{6 \,l^2}{N}\right)}{(N+1)(N+2)} -1 \right]\;\left(\|\mathbf{e}_l\|^2 - \| {\sf{fa}}_{l} \|^2 \right) \\
\hspace{0.4cm} \stackrel{\Delta}{=}\alpha_{N}(l)\; \left(\|\mathbf{e}_l\|^2 - \| {\sf{fa}}_{l} \|^2 \right)\;,\\
\sigma^{2}=\frac{ (N+1)^{2} (N+2)^{2}}{\left[ Q_1(l,N) + 2l\delta Q_2(l,N) + l^2 \delta^2 Q_3(l,N) \right] }\;{\|\mathbf{e}_{l}-{\sf{fa}}_{l} \|}^{2} \\
\hspace{0.5cm} \stackrel{\Delta}{=} \beta_{N}(l)\;{\|\mathbf{e}_{l}-{\sf{fa}}_{l} \|}^{2}\; .
\end{array}
\label{m-sigma}
\right.
\end{equation}
From eq. \ref{m-sigma}, we see that $\Delta_{f,c}(N)$ is normally distributed with an almost constant mean (roughly $\left(\|{\sf{fa}}_{l} \|^{2}-\|\mathbf{e}_l\|^2   \right)$), while its variance is proportional to $\sigma_{N}=\frac{1}{N} \;\|\mathbf{e}_{l}-{\sf{fa}}_{l} \|$, which will be of constant use from now. The situation is depicted in  fig. \ref{msigma}. In this figure, we see that $m$ is almost constant as $N$ increases, while its variance $\sigma$ increases. This results in an increase of $P(\Delta_{f,c}(N)\geq 0)$ since the darked area on the left of the $)$ threshold is decreasing.
This section will be divided in three subsections corresponding to the main steps of the development. The first idea consists in approximating the above normal density by a sum of indicator functions. Then, we have to calculate specific integrals (named $A_{i}$ and $B_{i}$ integrals). This will constitute the major difficulty since these integrals are defined on an implicitly defined domain.

\subsection{Approximating the normal density by a sum of indicator functions}
\label{approx-indic}

\begin{figure}[ht!]
\centering
\scalebox{0.43}{\includegraphics{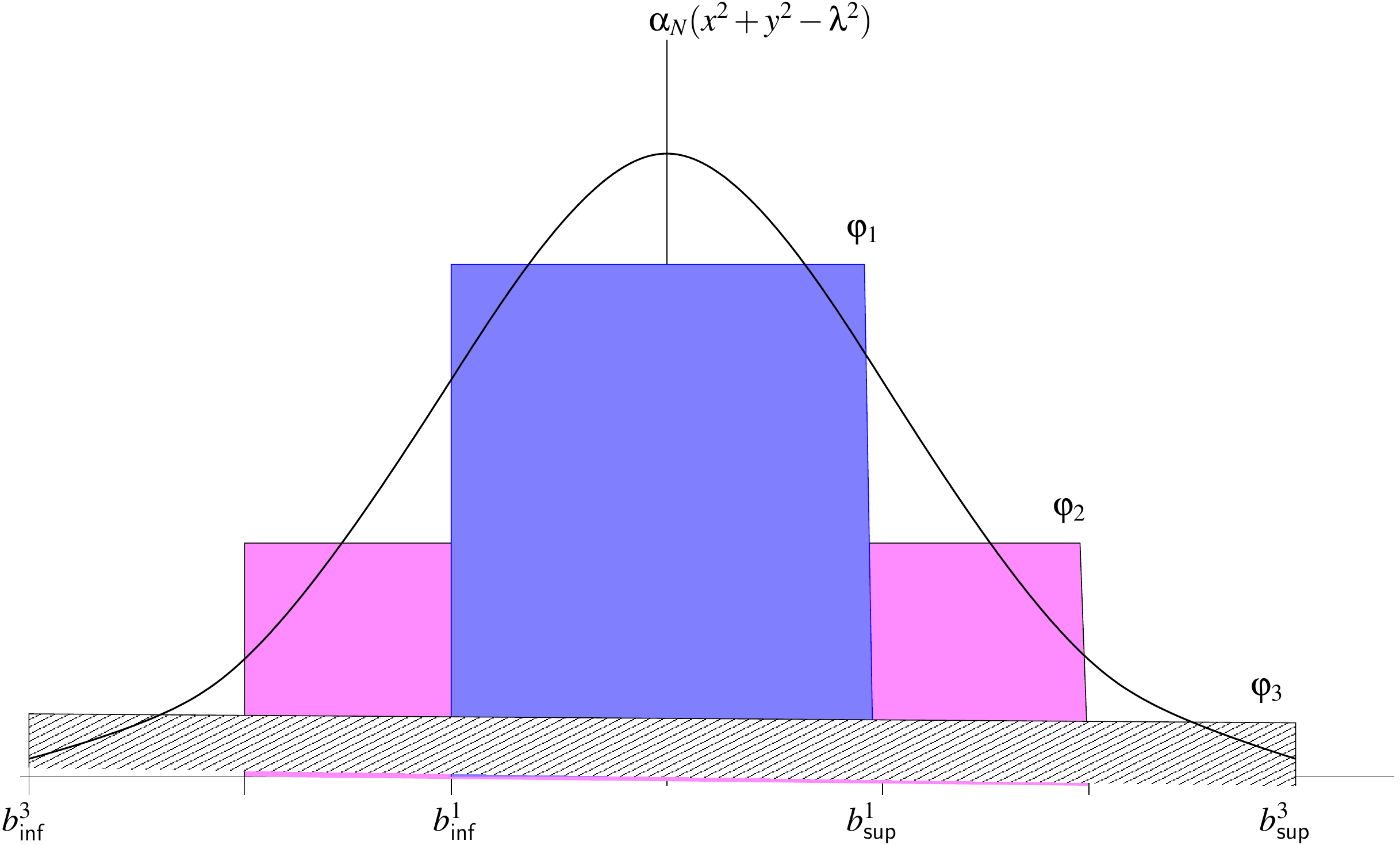}}
\caption{\it The  approximation scheme: the $\varphi_{i}$ functions }
\label{phi_i}
\end{figure}

A first step will consist in approximating  the density  $\mathcal{L}\left(\Delta_{f,c}\vert \,\tilde{\beps}_{l}= \mathbf{e}_{l} \right)$  (see eqs. \ref{normalun}, \ref{m-sigma}) by a weighted sum of $n$ indicator functions (denoted $\varphi_{i}$). Thus considering a "$3 \sigma$"\footnote{Of course, the choice of $3 \sigma$ is completely arbitrary  and extending our calculations to a $\kappa \sigma$ support is quite straightforward. Moreover, a $3 \sigma$ support is quite sufficient under the Gaussian assumption.} support of this approximation centered on the mean $m$ of this normal density, i.e. $\left[m-3 \sigma, m+3 \sigma \right]$ leads to:
\begin{equation}
\begin{array}{l}
\mathcal{L}\left(\Delta_{f,c}\vert \,\tilde{\beps}_{l}= \mathbf{e}_{l} \right)  \simeq \displaystyle{\sum_{i=1}^n}\; \frac{\gamma_i}{6\frac{i}{n} \;\sigma(x,y) } \; \varphi_{i}(x,y)\;,\\
\mbox{where:}\\
\varphi_{i}(x,y)\stackrel{\Delta}{=}\mathbf{1}_{ \Delta_{f,c}\in [b_{\sf{inf}}^{i}(x,y)\;,\;b_{\sf{sup}}^{i}(x,y)]}\;,\;\mathbf{e}_{l}=(x,y)^{T}\;.
\end{array}
\label{indic-1}
\end{equation}
This means that the supports of these $n$ indicator functions vary from  $[-3\,\frac{\sigma}{n}, 3\,\frac{\sigma}{n}]$, to $[-3\sigma, 3\sigma]$, and that we have the following definitions (see fig. \ref{phi_i}):
\begin{eqnarray}
\sigma(x,y) &=& 2 \sqrt{{( \mathbf{e}_{l}-{\sf{fa}}_{l}) }^T \Phi ( \mathbf{e}_{l}-{\sf{fa}}_{l}) } \;,\nonumber\\
{}&=& 2 \sqrt{\beta_N(l) \;\left[(x)^2+(y+\lambda)^{2} \;\right] } \;, \nonumber\\
b_{\sf{sup}}^{i}(x,y) &= & m(x,y) +3 \,\frac{i}{n}\;\sigma(x,y)\;,\nonumber \\
&=& {\sf{fa}}^T \cam {\sf{fa}} - (\mathbf{e}_{l})^T \cam (\mathbf{e}_{l})+\frac{3i}{n}\;\sigma(x,y)\;,\nonumber\\
{}&=& \alpha_N(l) (x^2+y^2-\lambda^2) + \frac{3i}{n}\; \sigma(x,y) \; \nonumber  \\ 
b_{\sf{inf}}^{i}(x,y)&=& {\sf{fa} }^T \cam {\sf{fa} }- (\mathbf{e}_{l})^T \cam (\mathbf{e}_{l}) - \frac{3i}{n}\; \sigma(x,y)\;,\nonumber\\
{}&=& \alpha_N(l) (x^2+y^2-\lambda^2) - \frac{3i}{n}\; \sigma(x,y)\;.
\label{bsup-binf}
\end{eqnarray}
The scalar parameters $\alpha_{N}(l)$ and $\beta_{N}(l)$ are given by (see eq. \ref{explicit_1}):
\begin{equation}
\left\{
\begin{array}{l}
\alpha_{N}(l) = [ \frac{2(2N+1-6l+6\frac{l^2}{N})}{(N+1)(N+2)} - 1 ]   \;,\\ \\
\beta_{N}(l) = \frac{Q_1(l,N)+2l \delta Q_2(l,N)+l^2 \delta^2 Q_3(l,N)}{(N+1)^2(N+2)^2} \;.
\end{array}
\right.
\label{alp-bet}
\end{equation}
For instance, for $l=N$, we have more simply :
\begin{equation}
\left\{
\begin{array}{l}
\medskip
\alpha_N =\frac{N(1-N)}{(N+1)(N+2)} \approx -1  \;, \nonumber \\ 
\beta_N = \frac{4N^3+226N^2-66N+4}{(N+1)^2(N+2)^2} \approx \frac{4}{N} \;(N\gg 1) \;.
\end{array}
\right.
\end{equation}
The fact that $\beta_{N}(l)$ is small (w.r.t. $1$) will play a central role for deriving closed form approximations of  $P \left(\Delta_{f,c}\geq 0 \right)$. The definition and meaning of the $\varphi_{i}$ functions are represented on fig. \ref{phi_i}.
With these definitions, we thus have the following approximation:
\begin{proposition}
\label{prop-1}
Consider the approximation of $P \left(\Delta_{f,c}\geq 0\vert \,\tilde{\beps}_{l}= \mathbf{e}_{l} \right)$ as a sum of indicator functions (see eq. \ref{indic-1}), the following equality holds true:
\begin{equation}
\label{eqprop-1}
\begin{array}{l}
P\left(\Delta_{f,c}\geq 0\vert \,\tilde{\beps}_{l}= \mathbf{e}_{l} \right) = \\
\displaystyle{\sum_{i=1}^{n}} \gamma_i \left[ \frac{b_{\sf{sup}}^{i}(x,y) } {2 \frac{3i}{n} {\sf{den}}(x,y)} \;{\mathbf{1}}_{b_{\sf{sup}}^{i}(x,y)\geq 0}-
\frac{b_{\sf{inf}}^{i}(x,y) }{2 \frac{3i}{n} {\sf{den}}(x,y)} \;{\mathbf{1}}_{b_{\sf{inf}}^{i}(x,y)\geq 0}\right] \;,\\ \\
=\displaystyle{\sum_{i=1}^{n}}\frac{\gamma_i}{2} \left( \mathbf{1}_{b_{\sf{sup}}^{i}(x,y)\geq 0}+ \mathbf{1}_{b_{\sf{inf}}^{i}(x,y)\geq 0}  \right)\\
+ \frac{ n}{12}\, \frac{\alpha_{N}}{\sqrt{\beta_N}}\;\frac{(x^2+y^2-\lambda^2)}{\sqrt{((x-\lambda)^2+y^2)}}\; \displaystyle{\sum_{i=1}^n} \frac{\gamma_i}{i}\;\left(\mathbf{1}_{b_{\sf{sup}}^{i}(x,y)\geq 0}- \mathbf{1}_{b_{\sf{inf}}^{i}(x,y)\geq 0}   \right) \;.
\end{array}
\end{equation} 
Moreover, we have:
\begin{equation}
b_{\sf{sup}}^{i}(x,y)\geq 0 \Longleftrightarrow f(x,y) \leq \frac{-6i}{n} \frac{\sqrt{\beta_N}}{\alpha_N}\;.
\label{eqprop-1bis}
\end{equation}
where:\\ 
\begin{tabular}{|c|}
\hline\\
$ f(x,y)=\frac{x^{2}+y^{2}-\lambda^{2}}{\sqrt{x+(y+\lambda)^{2}} }\;.  $
\\
\hline
\end{tabular}
\end{proposition}
{\bf Proof}: For the sake of completeness, a short proof is now presented. First, consider eq. \ref{eqprop-1} and assume that $\Delta(u) \in [b_{inf},\, b_{sup}]$. Then:
\begin{equation}
\begin{array}{lll}
\hspace{-0.5cm}\displaystyle{\int_{\Delta \geq 0} } {\mathbf{1}}_{\Delta(u) \in [b_{i}, b_{s}]}du &=& (b_{sup}-b_{inf}){\mathbf{1}}_{b_{inf} \geq 0} \\
&& + b_{sup}({\mathbf{1}}_{b_{sup} \geq 0}{\mathbf{1}}_{b_{inf} \geq 0}) \;,\\
\medskip
&=& b_{sup}\;\left({\mathbf{1}}_{b_{inf} \geq 0}+{\mathbf{1}}_{b_{sup} \geq 0}{\mathbf{1}}_{b_{inf} \leq 0} \right)\\
&& -b_{inf}\;{\mathbf{1}}_{b_{inf} \geq 0}\;, \\
\medskip
&=& b_{sup}\;\left( {\mathbf{1}}_{b_{sup} \geq 0} \underbrace{\left( {\mathbf{1}}_{b_{inf} \geq 0}+ {\mathbf{1}}_{b_{inf} \leq 0}\right)}_{\mathbf{1}}\,\right)\\
&&-b_{inf}\;{\mathbf{1}}_{b_{inf} \leq 0}\;, \\ 
&=& b_{sup}{\mathbf{1}}_{b_{sup} \geq 0} - b_{inf}{\mathbf{1}}_{b_{inf} \geq 0} \;.
\end{array}
\end{equation}
The first part of eq. \ref{eqprop-1} is thus proved. The second part of eq. \ref{eqprop-1} is a straightforward consequence of the expressions of $b_{\sf{sup}}^{i}(x,y)$ and $b_{\sf{inf}}^{i}(x,y)$ as given by eq. \ref{bsup-binf}.\\
The second part of Prop. \ref{prop-1} is also quite straightforward (notice that $\alpha_{N}(l)$ is negative):
\begin{equation}
\begin{array}{lll}
b_{sup}^{i}(x,y) \geq 0 & \iff & \alpha_N(x^2+y^2-\lambda^2) \\
&& + \frac{6i}{n}\sqrt{\beta_N}\sqrt{x^2 +(y+\lambda)^2} \geq 0 \;, \\ 
{} & \iff & f(x,y) \leq \frac{-6i}{n} \frac{\sqrt{\beta_N}}{\alpha_N} \;.
\end{array}
\end{equation}
$\Box$ $\Box$ $\Box$

The  $\left\{\gamma_{i}\right\}$ coefficients are obtained as the solution of an optimization problem (e.g. least squares, see Appendix B).
% At least they must satisfy to the following conditions:
%\begin{eqnarray}
%\sum_{i=1}^n \gamma_i & = & 1  \;, \nonumber \\
%\sum_{i=1}^n \frac{\gamma_i}{6\frac{i}{n}{\sf den}} & = & \frac{1}{{\sf den} \sqrt{2 \pi}} e^{0} \;.
%\end{eqnarray}
We stress that these $\left\{\gamma_{i}\right\}$ coefficients are considered as {\it fixed} whatever the value of the $\mathbf{e}_{l}$ vector. So, integrating over all the possible values of the $\mathbf{e}_{l}$ vector, we obtain:
\begin{eqnarray}
P(\Delta_{f,c} \geq 0) &=& \int_{\mathbb{R}^{2}} \;P \left(\Delta_{f,c} \geq 0 \vert \,\tilde{\beps}_{l}= \mathbf{e}_{l} \right) \;dx\,dy\;,\nonumber\\
 &=& \displaystyle{\sum_{i=1}^n}\frac{\gamma_{i}}{2} A_{i}+\frac{\alpha_{N}}{\sqrt{\beta_{N}} } \frac{n}{12} \displaystyle{\sum_{i=1}^{n}} \frac{\gamma_{i}}{i} \;B_{i}\;,
\label{pfc-int}
\end{eqnarray}
where:
\begin{equation}
\hspace{-1.2cm}\left\{
\begin{array}{l}
A_{i}=\displaystyle{\int_{\mathbb{R}^{2}} }\;\mathcal{N}_{(0,1)}(x,y) \left[ \mathbf{1}_{f(x,y)\leq -\frac{6 i\sqrt{\beta_{N}}}{n\,\alpha_{N}} }+\mathbf{1}_{f(x,y)\leq \frac{6 i\sqrt{\beta_{N}}}{n\,\alpha_{N}} } \;\right]\,dx dy,\\\\
B_{i}=\displaystyle{\int_{\mathbb{R}^{2}} }\mathcal{N}_{(0,1)}(x,y)f(x,y) \left[ \mathbf{1}_{f(x,y)\leq -\frac{6 i\sqrt{\beta_{N}}}{n\,\alpha_{N}} }-\mathbf{1}_{f(x,y)\leq \frac{6 i\sqrt{\beta_{N}}}{n\,\alpha_{N}} } \right]dx dy\;,
\end{array}
\right.
\end{equation}

For reasons which will clearly appear soon, it is worth to rewrite the $A_{i}$ and $B_{i}$ integrals as:
\begin{equation}
\left\{
\begin{array}{l}
B_{i} = \displaystyle{\int_{ \frac{6 i\sqrt{\beta_{N}}}{n\,\alpha_{N}} \leq f(x,y)\leq -\frac{6 i\sqrt{\beta_{N}}}{n\,\alpha_{N}} } }\;\mathcal{N}_{(0,1)}(x,y)\;f(x,y)\;dxdy\;,\\\\
A_{i} = \displaystyle{\int_{ \frac{6 i\sqrt{\beta_{N}}}{n\,\alpha_{N}} \leq f(x,y)\leq \frac{-6 i\sqrt{\beta_{N}}}{n\,\alpha_{N}}  } }\;\mathcal{N}_{(0,1)}(x,y)\;\;dxdy\;+\;\\
\hspace{2cm} 2\;\displaystyle{\int_{ f(x,y) \leq \frac{6 i\sqrt{\beta_{N}} }{n\,\alpha_{N}}  } }\;\mathcal{N}_{(0,1)}(x,y)\;\;dxdy\;.
\end{array}
\right.
\end{equation} 
So, now the problem we have to face is to obtain accurate closed form approximations of the $B_{i}$ and $A_{i}$ integrals.

\subsection{Approximating the $B_{i}$ integrals}
\label{integ-domains}

\begin{figure}[ht!]
\scalebox{0.35}{\includegraphics{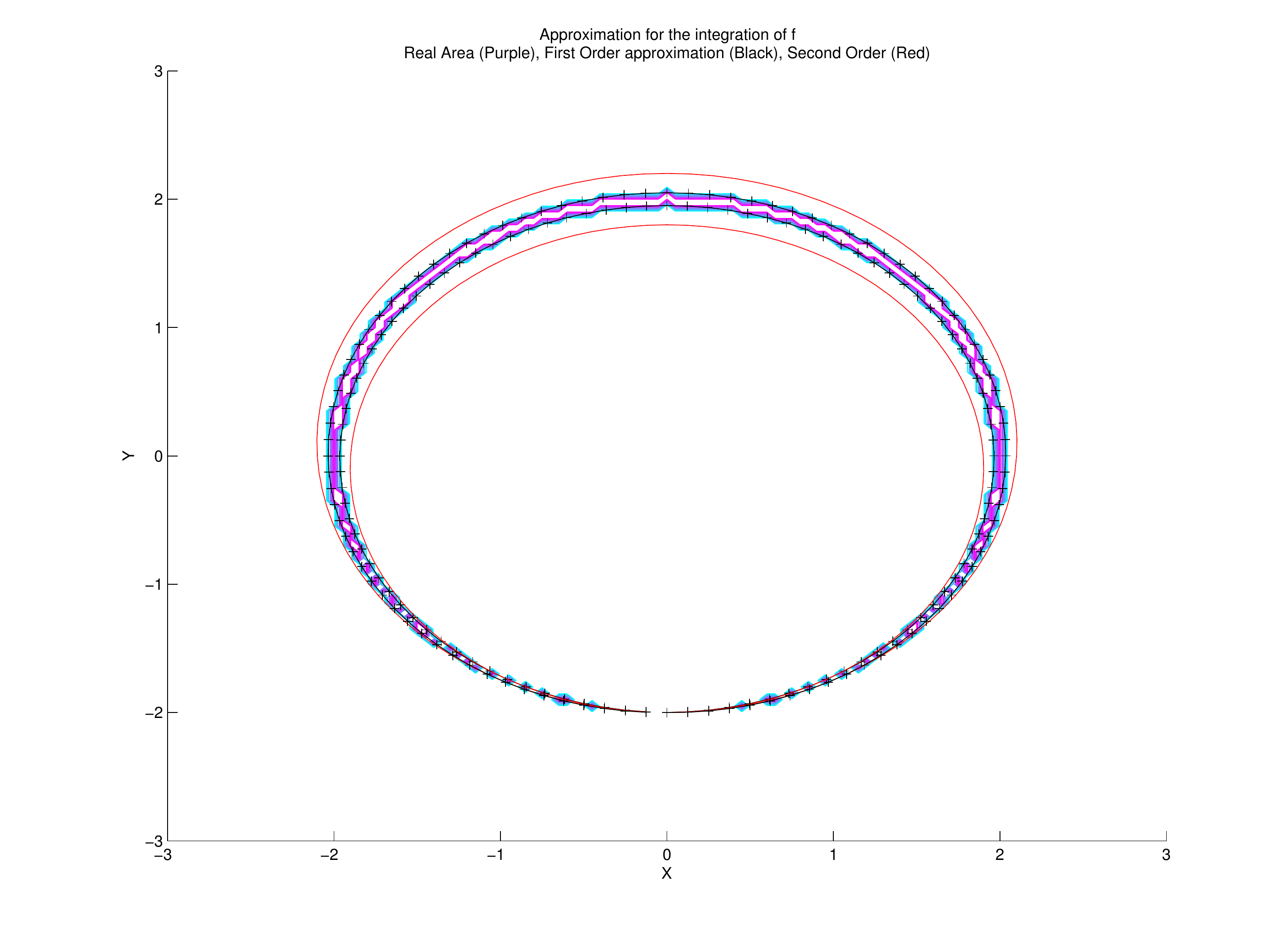}}
\caption{\it The  $f(x,y)$ function and its approximation (real: purple; approximations: continuous red and black)}
\label{graphcoeur}
\end{figure}

It is clear that deriving a general closed-form expression for the $B_{i}$ (or $A_{i}$) integrals is hopeless\footnote{There does not exist a primitive function of $\mathcal{N}_{(0,1)}(x,y)\;f(x,y)$ and the integral is {\bf implicitly} defined}. However, an accurate closed-form approximation can be obtained thanks to the following remark. When the scan number $N$ becomes great, then the ratio $\rho = \frac{\sqrt{\beta_N}}{\alpha_N}$ is close to zero. Now, the numerator of the $f(x,y)$ function  is zeroed on a circle (equation $x^{2}+y^{2}=\lambda^{2}$). This leads us to consider the following parametrization of the $(x,y)$-plane.
\begin{equation}
\left\{
\begin{array}{l}
x=(-\lambda+\varepsilon)\;\sin(\theta)\;,\\
y=(-\lambda+\varepsilon)\;\cos(\theta)\;.
\end{array}
\right.
\label{change_0}
\end{equation}
The function $f(x,y)$ is then changed in a $f(\varepsilon,\theta)$ function defined below, which leads to the following changes for the $B_{i}$ integral:
\begin{equation}
\left\{
\begin{array}{l}
f(\varepsilon, \theta)= \frac{-\varepsilon(2\lambda-\varepsilon)}{\sqrt{4\lambda \sin^2(\theta/2)(\lambda-\varepsilon)+\varepsilon^2}}\;\\
\exp\left(-\frac{x^2+y^2}{2}\right) =  \exp\left(-\frac{(\lambda-\varepsilon)^2}{2} \right)\;,\\
dxdy= \left| \lambda-\varepsilon \right| \;d\varepsilon \,d\theta\;.
\end{array}
\right.
\label{change_1}
\end{equation} 
Now, since we are considering only the small values of the $f$ function (numerator $(f) =-\varepsilon(2\lambda-\varepsilon)$), it is quite legitimate \footnote{Actually, there are two values of $\varepsilon$ zeroing  the  numerator of $f(\varepsilon,\theta)$, $\varepsilon=0$ and $\varepsilon=2 \lambda$. However, both are represented by a unique transformation (see eq. \ref{change_0}) }  to restrict our analysis to small values of $\varepsilon$. More precisely, we assume $\varepsilon\ll \lambda$. Then, the {\bf second} order expansion of the $f(\varepsilon,\theta)$ functional is :
\begin{equation}
f(\varepsilon, \theta) \stackrel{2}{=}\frac{-\varepsilon}{\left| \sin(\theta/2) \right|}\;.
\end{equation}
Practically, this is rather important since the integration domain which was previously implicitly defined is now {\bf explicitly} defined; i.e. it simply becomes:
\begin{equation}
\left\{
\begin{array}{l}
-\left| \sin(\theta/2) \right|\; \eta_{i,N}\leq \varepsilon \leq  \left| \sin(\theta/2) \right|\;\underbrace{\left(\frac{-6i\; \sqrt{\beta_N}}{n\;\alpha_N} \right)}_{\eta_{i,N}}\;,\\
0\leq \frac{\theta}{2} \leq  \pi\;.
\end{array}
\right.
\label{cardio1}
\end{equation} 
The accuracy of this approximation is illustrated by fig. \ref{graphcoeur}. We can notice that the integration domain is well approximated.
The integration having been conveniently approximated, we consider also a second order expansion of the integrand $F(\varepsilon,\theta)$ of the $B_{i}$ integral, i.e. with:
$$
F(\varepsilon,\theta)=f(\varepsilon, \theta)\mathcal{N}(\varepsilon, \theta) \left|J(\varepsilon, \theta)\right|\;,
$$
and $\left|J(\varepsilon, \theta)\right|=|\lambda-\varepsilon|$ the Jacobian of the $(x,y)\rightarrow (\varepsilon, \theta)$ transform, we have:
\begin{equation}
F(\varepsilon,\theta)\stackrel{2}{=}-\lambda \varepsilon \;\frac{e^{-\lambda^2/2}}{\left| \sin(\theta/2) \right|}+\frac{(1-2 \lambda^2)}{2 \left| \sin(\theta/2)\right|} e^{-\lambda^2/2} \;\varepsilon^2 \;.
\end{equation}
Considering on the first hand the effect of changing $\varepsilon$ into $-\varepsilon$ for this $2$-nd order expansion and the integration domain on the second one,  the effect of the $\varepsilon$ term is zero, so that:
\begin{equation}
\begin{array}{lll}
B_i &=& \frac{1}{2\pi}\;\displaystyle{\int_{\theta} \int_{\varepsilon=-\eta_{i,N}\,\sin(\theta/2) }^{\varepsilon=\eta_{i,N} \,\sin(\theta/2) } } \;\;\frac{(1-2\,{\lambda}^2)}{2\left| \sin(\theta/2) \right|} \;e^{-\frac{(\lambda)^2}{2}}\;\varepsilon^2 \; d\eps\;d\theta\;, \\\\
&=& \frac{(1-2\lambda^2)}{2\pi}\;e^{-\frac{(\lambda)^2}{2}} \;\frac{\eta_{i,N}^3}{3}\;\displaystyle{\int_{\theta} } \left(\sin(\theta/2)\right)^{2}d\theta\;,
\end{array}
\label{Bifin}
\end{equation} 
where $\eta_{i,N}= \frac{-6\,i}{n} \;\frac{ \sqrt{\beta_N}}{\alpha_N}$ (see eq. \ref{cardio1}).  Thus, a very simple closed-form approximation of the $B_{i}$ integral has been obtained, from which the following approximation of the part $\frac{\alpha_N}{\sqrt{\beta_N}} \frac{n}{12} \displaystyle{\sum_{i=1}^{n}} \frac{\gamma_i}{i}B_i$ of $P(\Delta_{f,c} \geq 0)$ (see eq. \ref{pfc-int}) is deduced:
~\\
\begin{tabular}{|c|}
\hline\\

$
\frac{\alpha_N}{\sqrt{\beta_N}} \;\frac{n}{12} \displaystyle{\sum_{i=1}^{n}} \frac{\gamma_i}{i}B_i \simeq 3(1-2\lambda^2)e^{-\lambda^2/2}\;\frac{\beta_N}{\alpha_N^2}\; \left(\frac{\displaystyle{\sum_{i=1}^{n} } i^2\;\gamma_i}{32 \;n^2} \right)\;.

$
\\
\hline
\end{tabular}
~\\
Thus, we see that an accurate approximation of the term $\frac{\alpha_N}{\sqrt{\beta_N}} \frac{n}{12} \displaystyle{\sum_{i=1}^{n}} \frac{\gamma_i}{i}B_i$ is proportional both to the ratio $\frac{\beta_N}{\alpha_N^2} \propto \frac{1}{N}$ and the fixed term  $3(1-2\,\lambda^{2})e^{-\lambda^2/2}$.

\subsection{Approximating the $A_i$ integrals}

We have now to turn toward the  $A_i$ terms. First, we remark that:

\begin{equation}
\begin{array}{l}
{\bf{1}}_{f(x,y)\leq -\eta_{i,N}}+{\bf{1}}_{f(x,y) \leq \eta_{i,N}}={\bf{1}}_{-\eta_{i,N} \leq f(x,y) \leq \eta_{i,N}} \\
+ 2\left( {\bf{1}}_{f(x,y)\leq 0}-{\bf{1}}_{-\eta_{i,N} \leq f(x,y) \leq 0} \right)\;,
\end{array}
\end{equation} 
so that, we have:
\begin{equation}
\begin{array}{lll}
A_i &=&  2\,\underbrace{\int_{\mathbb{R}^2} \mathcal{N}_{(0,1)}(x,y)\; ({\bf{1}}_{f(x,y)\leq 0}-{\bf{1}}_{-\eta_{i,N} \leq f(x,y) \leq 0}) dxdy}_{A_{i,1}} \;,\nonumber\\ \\
&+&\underbrace{\int_{\mathbb{R}^2} \mathcal{N}_{}(x,y) {\bf{1}}_{-\eta_{i,N} \leq f(x,y) \leq \eta_{i,N}} dxdy}_{A_{i,2}}\;.
\end{array}
\label{ai1}
\end{equation} 
We use the same change of variable (see eq. \ref{change_1}) as previously. For the $A_{i,1}$ integral the normal density is integrated over the $(\varepsilon,\theta)$ domain $[0,2\lambda]\times[0,2 \pi]$; while for the $A_{i,2}$ integral it is $[0,\eta_{i,N}\left|\sin(\theta/2)\right|]\times[0,2\pi]$. We thus have:

\begin{equation}
\begin{array}{lll}
\medskip
A_{i,1}&=& \frac{1}{\pi} \displaystyle{\int_0^{2\pi}} [e^{-(\lambda-\varepsilon)^2/2}]_{0}^{\lambda} -[e^{-(\lambda-\varepsilon)^2/2}]_{\lambda}^{2\lambda} d\theta \;,\\
\medskip
& & + \frac{1}{ \pi} \displaystyle{\int_0^{\pi}} [e^{-(\lambda-\varepsilon)^2/2}]_{0}^{\eta_{i,N} \left| \sin(\theta/2) \right|} d\theta \; ,\\
&\simeq& 2-e^{-\lambda^2/2}\;[2+2\,\eta_{i,N} - \frac{(\lambda^2-1)}{4}\,\eta_{i,N}^2]
\end{array}
\end{equation} 
For the $A_{i,2}$ integral, we proceed in the same way that for $B_i$, i.e. :

\begin{equation}
\begin{array}{lll}
A_{i,2} &=& \frac{1}{2\pi} \displaystyle{\int_0^{2\pi}} \left[e^{-\frac{(\lambda-\eta_{i,N} \;\left| \sin(\theta/2) \right|)^2}{2}} - e^{-\frac{(\lambda+\eta_{i,N} \;\left| \sin(\theta/2) \right|)^2}{2}} \right]d\theta\;,\\\\
&\simeq & \frac{2\lambda e^{-\lambda^2/2}}{\pi} \eta_{i,N} \;.
\end{array}
\end{equation}
Gathering the above results, we have just obtained a closed form approximation of the $A_{i}$ term:

\begin{equation}
A_i =\left(\frac{-2\pi+(2\lambda-2\pi)\eta_{i,N}+\frac{\pi}{4}(\lambda^2-1)\eta_{i,N}^2}{\pi}\,\right)\;e^{-\lambda^2/2}\;.
\end{equation} 

\subsection{The closed-form approximations of $P(\Delta_{f,c} \geq 0)$}

Summarizing the previous calculations, we are now in position to present the following result, which constitutes also the principal result of this paper.
\begin{proposition}
 Let us consider that the possible false association can occur at unique time period (denoted $l$), then  a closed-form approximation of the probability of correct association is:
\begin{center}
\begin{tabular}{|c|}
\hline\\
$P(\Delta_{f,c} \geq 0) = 1+(a+b\,\lambda+c\,\lambda^{2})\;e^{-\frac{{\lambda}^{2}} {2} }$
\\
\hline
\end{tabular}
\end{center}
where:
\begin{equation}
\left\{
\begin{array}{lll}
\medskip
a&=&-\frac {1} {2 \pi} \;\left[1+ \frac{\sqrt{\beta_N(l)}}{\alpha_{N}(l) }\;\displaystyle{\sum_{i=1}^{n}}\frac{\gamma_{i} }{i}  +\frac{66\pi}{32 n^{2}}\,\frac{\beta_{N}(l)}{\alpha_{N}^{2}(l)}\;\displaystyle{\sum_{i=1}^{n}} i^{2}\, \gamma_{i} \,\right]  , \\ 
\medskip
b&=& \frac{1} {2 \pi} \;\left[\frac{6}{n}\frac{\sqrt{\beta_N}(l)}{\alpha_N(l)} \; \displaystyle{\sum_{i=1}^{n}} i\;\gamma_{i} \,\right] ,\\ 
c&=& \frac{15}{16 n^{2}}\,\frac{\beta_N(l)}{\alpha_N^{2}(l)} \; \displaystyle{\sum_{i=1}^{n}} i^{2}\,\gamma_{i} \;.
\end{array}
\right.
\label{machine}
\end{equation} 
The scalars $\alpha_{N}(l)$ and $\beta_{N}(l)$ are given by eq. \ref{alp-bet}.
\label{prop_2}
\end{proposition}
This formula is quite simple and relevant. We can notice also that $P(\Delta_{f,c} \geq 0)$ is independent of the kinematic scenario parameters, since it involves only the ratio $\lambda/ \sigma$ ({\small here simply denoted $\lambda$}), and the  number of scans $N$ (via $\alpha_{N}(l)$ and $\beta_{N}(l)$). \\
Since we have $\beta_N \propto \frac{1}{N}$ and $\alpha_{N} \propto -1$, the asymptotic value of $P(\Delta_{f,c} \geq 0)$ is simply $1-\frac{e^{-\frac{{\lambda}^{2}} {2} } }{2\;\pi}$. This rough approximation is valid for values of $N$ as small as $30-40$. Not surprisingly, we see that the dimensioning parameter for $P(\Delta_{f,c} \geq 0)$ is the ratio $\lambda/ \sigma$.\\

Since  $\beta_N$ is small, it is the elementary increment. So, the slope (denoted {\sf{slo}}) of $P(\Delta_{f,c} \geq 0)$ as a function of $N$ is the factor \footnote{The superscript $f^{'}$ denoting the derivative, $\sqrt{x}^{'}=\frac{1}{2 \sqrt{x}}$ while $x^{'}=1$} of the ratio $\frac{\sqrt{\beta_N(l)}}{\alpha_{N}(l) }$, i.e. it is:
\begin{equation}
\begin{array}{l}
{\sf{slo}}=\frac{1}{2 \pi}\;\left(\frac{6}{n}\;\displaystyle{\sum_{i=1}^{n}} i\;\gamma_{i} -\displaystyle{\sum_{i=1}^{n}}\frac{\gamma_{i} }{i} \right)\;.\\
\mbox{so that :}\\
P(\Delta_{f,c} \geq 0) \stackrel{1}{\simeq} 1-\left(1-{\sf{slo}} \;\frac{\sqrt{\beta_N(l)}}{\alpha_{N}(l) } \right)\;e^{-\frac{{\lambda}^{2}} {2} }\;.
\end{array}
\label{machine1}
\end{equation}
Note that, for $N$ "great" ($30-40$) the approximation given by eq. \ref{machine1} is less precise that the approximation given by eq. \ref{machine}. However, its main interest is to put in evidence the effect of the $N$ parameter. If the $\left\{\gamma_{i} \right\}$ coefficients are determined by minimizing a least square criterion, then {\sf{slo}} can be easily calculated (see Appendix B), and is obviously positive (see eq. \ref{slo-g}).

\subsection{The case of a random $\lambda$}
\label{random_lambda}

Up to now, it was assumed that the parameter $\lambda$ was deterministic. However, it is more realistic to model this seducing measurement by a normal density $\mathcal{N}(\lambda_{0}, \sigma_{0})$. Let $\bar{\Delta}_{f,c}$ be the (extended) cost difference  for this $\lambda$ modeling, conditioning on $\lambda$, we then have:
\begin{equation}
\begin{array}{lll}
P(\bar{\Delta}_{f,c}\geq 0) &=& \mathbb{E}_{\lambda}\left[ P_{\lambda}(\Delta_{f,c}\geq 0) |\lambda \right]\;,\\
\mbox{with:}\\
P_{\lambda}(\Delta_{f,c} \geq 0) &=& 1+(a+b\lambda+c\lambda^2)e^{-\lambda^2/2}\;.
\end{array}
\end{equation} 
Performing straightforward calculations, we obtain:
\begin{equation}
\begin{array}{l}
P(\bar{\Delta}_{f,c}\geq 0) = 1+ \frac{1}{\sqrt{\sigma_0^2+1}}\;\left[ a+b\bar{\lambda}_0 + c (\bar{\lambda}^2_0+s_0^2) \right] e^{-\frac{\lambda_0^2}{2(\sigma_0^2+1)}}\;,\\
\mbox{where:}\\
{\bar{\lambda}}_{0} = \frac{1}{\sigma^2_0+1} \lambda_0 \qquad , \qquad s_0^2=\frac{\sigma_0^2}{\sigma_0^2+1} \;.
\end{array}
\end{equation} 
So, for $N$ sufficiently large, we have $P(\bar{\Delta}_{f,c}\geq 0) \approx 1-\frac{1}{\sqrt{\sigma_0^2+1}}\;\frac{e^{-\frac{{\bar{\lambda}_{0}}^{2}} {2} } }{2\;\pi}$. Thus, we see that the effect of this randomization of $\lambda$ is far to be negligible.

\subsection{A system analysis perspective}
Using the previous results, we are now turning our effort toward the steady-state behavior of the association process via a Discrete Time Markov Chain (DTMC) analysis. We consider that at each time period there is a binary decision process, defined by:
\begin{equation}
\left\{
\begin{array}{l}
\left[ \ca \right]\;:\mbox{event: correct association}\;,\; \left[\fa \right]\;:\mbox{event: false association}\;,\\
p_{\fa }\stackrel{\Delta}{=}\mbox{probability of false association}.
\end{array}
\right.
\end{equation}
Note that closed form approximations $p_{\fa}$ have been already obtained. We assume furthermore that $p_{\ca}=1-p_{\fa}$ and that this decision process can be modeled by an {\bf homogeneous} DTMC. We are interested now in the evaluation of the probability that $k$ consecutive false associations occur. We shall   focus on the case $k=2$. To that aim, let us define the  random variable $X$ which can take $4$ states, defined by:
\begin{equation}
\left|
\begin{array}{l}
\mbox{state: }(1): \left[\ca,\ca \right]\;\;,\;\;\mbox{state: }(2): \left[\ca,\fa \right]\;,\\
\mbox{state: }(3): \left[\fa,\ca \right]\;\;,\;\;\mbox{state: }(4): \left[\fa,\fa \right]\;.
\end{array}
\right.
\end{equation}
It is easily shown that $X$ is also a DTMC, whose  transition matrix (denoted ${\mathsf{P}}_{2}$) stands as follows:
\begin{equation}
{\mathsf{P}}_{2}=\left(
\begin{array}{llll}
 1-p_{\fa} & p_{\fa} & 0 & 0 \\
0 & 0 & 1-p_{\fa} &p_{\fa} \\
1-p_{\fa} & p_{\fa} & 0 & 0 \\
0 & 0 & 1-p_{\fa} &p_{\fa}
\end{array}
\right)
\end{equation}

Considering the transition matrix ${\mathsf{P}}_{2}$, we see that this DTMC  is aperiodic and irreducible, ensuring the existence of a stationary distribution \cite{kul}. State $4$ is especially relevant for our analysis, since it corresponds to {\bf two} consecutive false associations.
The structure of the matrix ${\mathsf{P}_{2}}^{2}$ is quite enlightening and is a characteristic feature. Indeed, straightforward calculations yield:
\begin{equation}
\begin{array}{l}
\medskip
{\mathsf {P}_{2}}^{2}=\left[{(1-p_{\fa})}^{2} \,\ind, p_{\fa}(1-p_{\fa}) \,\ind, p_{\fa}(1-p_{\fa}) \,\ind, p_{\fa}^{2} \,\ind \right]\;,\\
\mbox{where:}\; \ind\stackrel{\Delta}{=}{\left(1,1,1,1\right)}^{T}\;.
\end{array}
\end{equation}
 Thus, ${\mathsf {P}_{2}}^{2}$ admits the following factorization:
\begin{equation}
\begin{array}{l}
{\mathsf {P}_{2}}^{2}=\mathsf{V}\;{\mathsf{W}}^{T}\;,\\
\mbox{where:}\\
\mathsf{V}=\left(1-p_{\fa}\right) \, \ind \;,\;{\mathsf{W}}^{T}=\left(1-p_{\fa},\;p_{\fa},\;p_{\fa},\frac{p_{\fa}^{2}}{(1-p_{\fa}\;)} \right)\;.
\end{array}
\end{equation}
Furthermore, it is easily shown that ${\mathsf{W}}^{T}\,\mathsf{P}_{2}=\mathsf{W}^{T}$. Thus, we have:
\begin{eqnarray}
{\mathsf {P}_{2}}^{3}&=&\left(\mathsf{V}\;{\mathsf{W}}^{T}\right)\;\mathsf {P}_{2}\;,\\ \nonumber
&=&\mathsf{V}\;\left({\mathsf{W}}^{T}\,\mathsf{P}_{2} \right)\;,\\ \nonumber
&=&\left(\mathsf{V}\;{\mathsf{W}}^{T}\right)={\mathsf {P}_{2}}^{2}\;.
\end{eqnarray}
And more generally, whatever $n\geq 4$ we have ${\mathsf {P}_{2}}^{n}={\mathsf {P}_{2}}^{2}\;{\mathsf {P}_{2}}^{n-2}={\mathsf {P}_{2}}^{4}={\mathsf {P}_{2}}^{2}$, yielding the following result:
\begin{proposition}
Whatever $n\geq 2$, the following equality holds true:\\
 ${\mathsf {P}_{2}}^{n}={\mathsf {P}_{2}}^{2}$.
\end{proposition}
So, whatever the initial distribution $\mathsf{X}_{0}$, described by the {\bf row} vector $\mathsf{X}_{0} =\left(x_{1},x_{2},x_{3},x_{4}\right)$, we have ($\forall n \geq 2$):
\begin{eqnarray}
{\mathsf{X}_{0}}^{(n)} &=&\mathsf{X}_{0}\,{\mathsf {P}_{2}}^{n}=\mathsf{X}_{0}\,{\mathsf {P}_{2}}^{2}\;,\\ \nonumber
&=& \left(\mathsf{X}_{0}\,\mathsf{V} \right)\,{\mathsf{W}}^{T}\;,\\ \nonumber
&=&\left(1-p_{\fa}\right) \;\underbrace{\left(\mathsf{X}_{0}\,\ind \right)}_{=1}\,{\mathsf{W}}^{T}=\left(1-p_{\fa}\right)\,{\mathsf{W}}^{T}\;,\\ \nonumber
 &=& \left( \;(1-p_{\fa})^{2} \;, p_{\fa}\,(1-p_{\fa}) \;, p_{\fa}\,(1-p_{\fa})\;,p_{\fa}^{2} \right)\;.
 \end{eqnarray}
 Similarly, let us consider the (asymptotic) stationary distribution $\boldsymbol{\pi}$, then $\boldsymbol{\pi}$ is a solution of the balance equation 
$\boldsymbol{\pi}=\boldsymbol{\pi}\;{\mathsf{P}}_{2}$. Not surprisingly, it is easily shown that:
\begin{equation}
\boldsymbol{\pi}= \left( (1-p_{\fa})^{2} \;, p_{\fa}\,(1-p_{\fa}) \;, p_{\fa}\,(1-p_{\fa})\;, p_{\fa}^{2} \right)\;.
\end{equation}
 We are now in position for studying the behavior of this DTMC. Since the state $4$ is particularly important, let us recall the following classical result \cite{kul}, \cite{grim}.
\begin{proposition}
  Assume the DTMC is irreducible and let $\boldsymbol{\pi}$ its stationary distribution, then the mean inter-visit time $m_{j,j}$ is given by
$$
m_{j,j}=\frac{1 }{\pi_{j}}\;,\;1\leq j \leq N\;.
$$
\end{proposition}
 Thus, we have here $m_{4,4}=\frac{1}{\pi_{4}}=\frac{1}{p_{\fa}^{2}}$, a value which is usually very weak if $p_{\fa}$ is small. Consider now a slight modification of the DTMC. If the state $4$ is attained ,  then the DTMC {\it remains} on (the absorbing) state $4$. The associated transition matrix  $\tilde{\mathsf{P}}_{2}$ reads:
\begin{equation}
\tilde{\mathsf{P}}_{2}=\left(
\begin{array}{llll}
 1-p_{\fa} & p_{\fa} & 0 & 0 \\
0 & 0 & 1-p_{\fa} &p_{\fa} \\
1-p_{\fa} & p_{\fa} & 0 & 0 \\
0 & 0 & 0 & 1
\end{array}
\right)
\end{equation}
The aim of this modeling is to investigate the probability that the system  be {\bf at least} one time in state $4$, during a given time interval. To this aim, calculations are greatly simplified if the following rewriting of the $\tilde{\mathsf{P}}_{2}$ matrix is considered:
\begin{equation}
\tilde{\mathsf{P}}_{2}=\left(
\begin{array}{ll}
\mathsf{Q} & {\bf v}_{1}\\
{\bf 0}^{T} & 1
\end{array} \right) \;,
\end{equation}
where $\mathsf{Q}$ is a $3\times 3$ left-up matrix. Elementary calculations  yield:
\begin{equation}
{\tilde{\mathsf{P}}_{2}}^{n}=\left(
\begin{array}{ll}
{\mathsf{Q}}^{n} & {\bf v}_{n}\\
{\bf 0}^{T} & 1
\end{array} \right) \;.
\label{bruno1}
\end{equation}
If we are able to provide an explicit expression of ${\mathsf{Q}}^{n}$, there is no need to calculate the vector ${\bf v}_{n}$ since the matrix ${\tilde{\mathsf{P}}_{2}}^{n}$ is stochastic. The eigensystem of the  $Q$ matrix is quite simple, i.e. :
\begin{equation}
\left|
\begin{array}{ll}
\mbox{eigenvalues} &\mbox {eigenvectors} \\
\lambda_{1}=0 &  {\bf u}_{1}^{T}=\left(\frac{-p_{\fa}}{1-p_{\fa}},1,0\right)\\
\lambda_{2}=\frac{1}{2}\left(1-p_{\fa}-\sqrt{\delta}\right) &  {\bf u}_{2}^{T}=\left(1,\frac{(1-p_{\fa})}{\lambda_{2}},1\right)\\
\lambda_{3}=\frac{1}{2}\left(1-p_{\fa}+\sqrt{\delta}\right)  &  {\bf u}_{3}^{T}=\left(1,\frac{(1-p_{\fa})}{\lambda_{3}},1\right) \\
\delta=\left(1+2\;p_{\fa}-3 \;p_{\fa}^{2}\right)  &\;.
\end{array}
\right.
\end{equation}
From which the following equality is deduced\footnote{after normalization of the ${\bf u}_{2}$ and ${\bf u}_{3}$ vectors}:
\begin{equation}
Q^{n}=\lambda_{2}^{n}\;\left({\bf u}_{2} {\bf u}_{2}^{T}\right)+\lambda_{3}^{n}\;\left({\bf u}_{3} {\bf u}_{3}^{T}\right)\;.
\end{equation}
Consequently, admitting an initial distribution $\mathsf{X}=\left(1,0,0,0\right)$ of the system state, the probability that the state $4$ has been attained {\it at least} at one time within the temporal interval $\left[0,n\right]$ is:
\begin{equation}
{\tilde{\mathsf{P}}_{2}}^{n}\;(1,4)  = 1-\lambda_{2}^{n+1}\;\left(\frac{2\,\lambda_{2}+1-p_{\fa}}{2\,\lambda_{2}^{2}+{(1-p_{\fa})}^{2} } \right) +\lambda_{3}^{n+1}\;\left(\frac{2\,\lambda_{3}+1-p_{\fa}}{2\,\lambda_{3}^{2}+{(1-p_{\fa})}^{2} }\right)\;.
\end{equation}
A second order expansion (w.r.t. $p_{\fa}$)  gives us ${\tilde{\mathsf{P}}_{2}}^{n}\;(1,4) \simeq (n+1)p_{\fa}^{2}+\frac{p_{\fa}}{3}$.
To complete this analysis, let us denote $N_{a}$ the number of visits to the transient states, before visiting the absorbing state (state $4$ here), then we have:
\begin{equation}
P(N_{a}=n)={\mathsf{X}}_{0}^{T}\;{\mathsf{Q}}^{n-1}\;(\mathsf{Id}-\mathsf{Q})\;\ind\;,\;\;n\geq 1.
\end{equation}
Hence, the {\it expected} number of visits to the absorbing state is simply:
\begin{equation}
\begin{array}{l}
\mathbb{E}(N_{a})=\displaystyle{{\sum}_{n\geq 1} }\;n\;P(N_{a}=n)={\mathsf{X}}_{0}^{T}\;{(\mathsf{Id}-\mathsf{Q})}^{-1}\;\ind\;,\\
\mbox{with:}\\
{(\mathsf{Id}-\mathsf{Q})}^{-1}\;\ind=\left(\frac{1+p_{\fa}}{p_{\fa}^{2}},\frac{1}{p_{\fa}^{2}},\frac{1+p_{\fa}}{p_{\fa}^{2}} \right)\;.
\end{array}
\end{equation}
As $p_{\fa}$ is rather small for our application, we thus have $\mathbb{E}(N_{a})\simeq \frac{1}{p_{\fa}^{2}}$, whatever the initial distribution of the transient states. Extending the previous analysis to an arbitrary value of $k$ is straightforward and we simply refer to \cite{vanp}.\\
The advantage of this analysis is its simplicity. However, a strong assumption is that the $p_{\fa}$ at time $t+1$ is not modified if a false association has occurred at time $t$. If $k$ and the $p_{\fa}$ are sufficiently small, this is a realistic assumption. If a large number of consecutive false associations occurs the parameters of the regression are changed and we have to turn to a more precise approach. This will be the aim of section \ref{mfmc}.

\section{Simulation Results (unique false association)}

\begin{figure}[ht!]
\centering
\scalebox{0.45}{\includegraphics{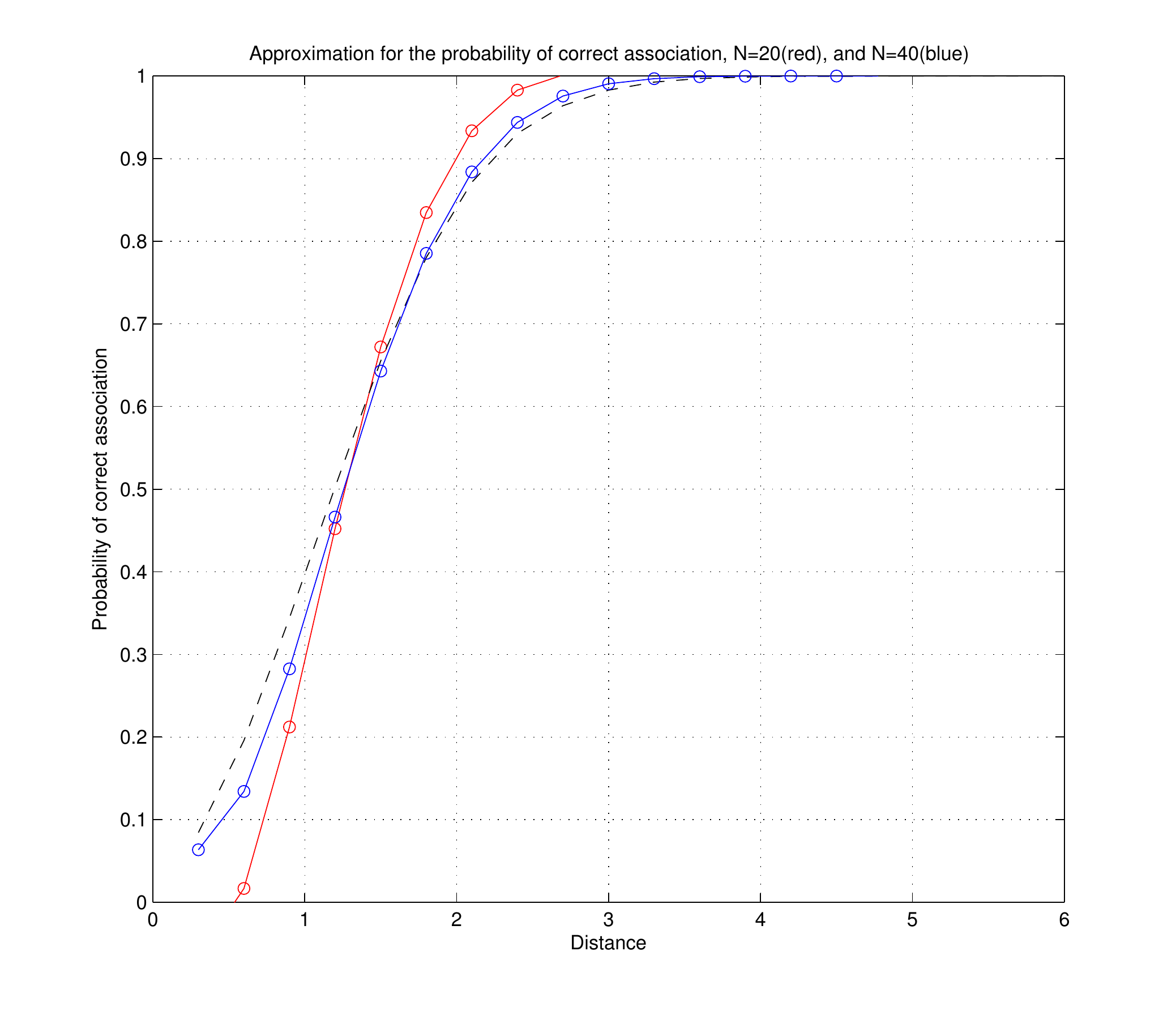}}
\caption{\it The  probability of correct association (dashed) $P({\Delta}_{f,c}\geq 0)$ and approximated (in red: $N=20$, in blue $N=40$), versus $\lambda$ ($x-axis$).}
\label{Probalambda}
\end{figure}

Once we have get the main result (eq. \ref{machine}), we have to test the accuracy of  our approximations. For doing that, we just have to consider the variations of the two dimensioning parameters ($\lambda$ and $N$). For the first one ($\lambda$), the number of scans ($N$) is a fixed value ($N=20$ and $N=40$). Then, we compare the exact value of $P(\Delta_{f,c} \geq 0)$ and its approximation as given by eq. \ref{machine}, for increasing values of the $\lambda$ parameter. Note that $\lambda$ represents in fact the ratio $\lambda/\sigma$ where $\lambda$ is the distance between the exact target position and the position of the "false" target, while $\sigma$ is the observation noise standard deviation. The result is displayed on fig. \ref{Probalambda}.
We can see that our approximation (eq. \ref{machine}) performs quite satisfactorily in general,  but is better as N increases. This is not surprising, especially if we remind that our approximations were based on the fact that the integration bounds $\eta_{i,N}$ were small, meaning that $N$ was sufficiently  great. \\
 This approximation is valid for value of $\lambda$ as small as $1$, which has only a mathematical meaning since for this value of $\lambda$ it is quite likely that measurements are merged. A complete derivation of the probability density function (pdf) of merged measurements has been performed in \cite{chanb}, \cite{ng}. However, it seems hopeless to include unresolved measurement pdf in our calculations for a closed form approximation of $P(\Delta_{f,c} \geq 0)$.
We can see that for $\lambda$ values between $1$ and $2$, the slope of $P({\Delta}_{f,c}\geq 0)(\lambda)$ is almost constant and rather important.  When $\lambda$ becomes close to $3$, then the probability of correct association is very close to $1$. \\

\begin{figure}[ht!]
\centering
\scalebox{0.60}{\includegraphics{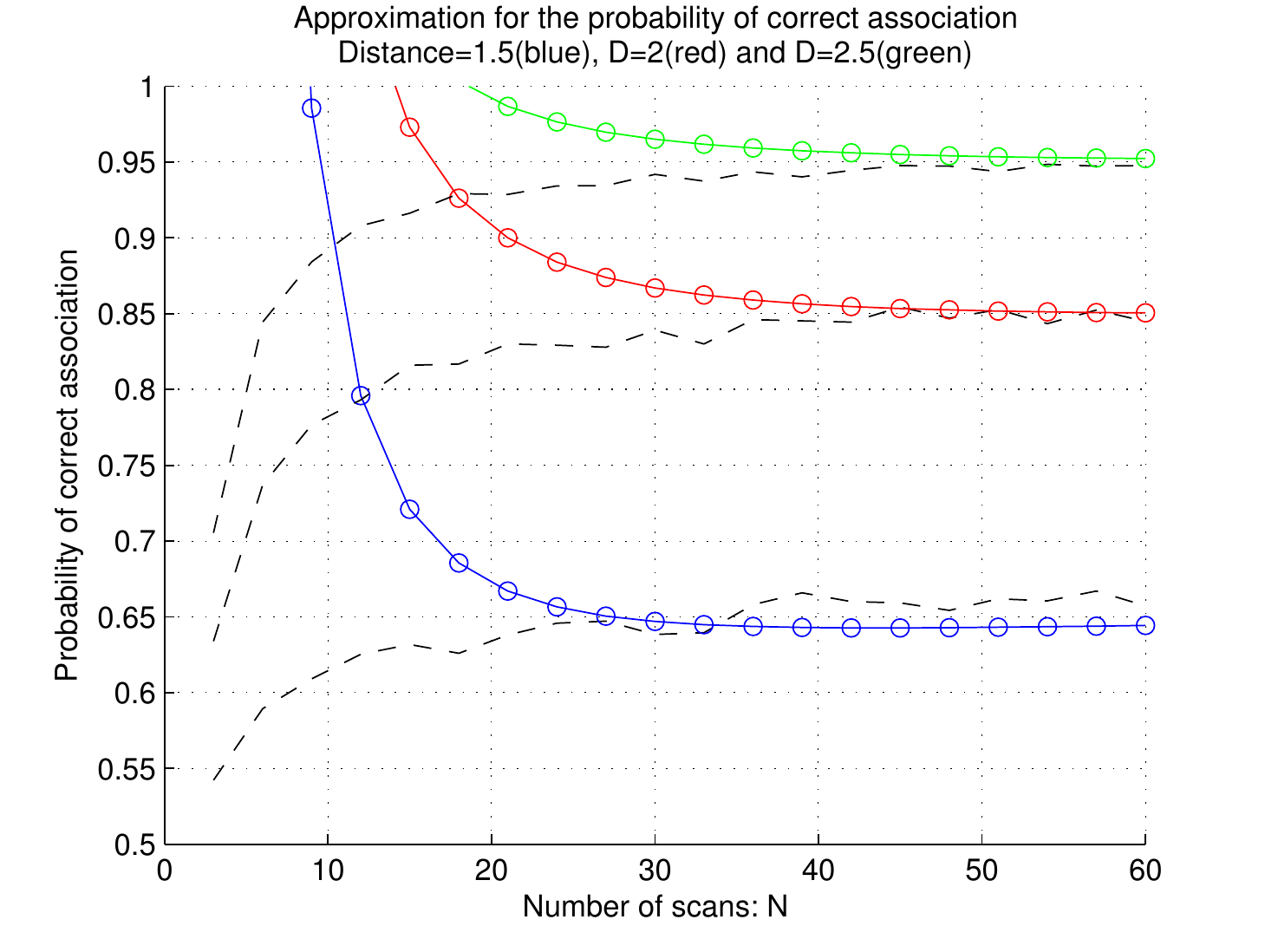}}
\caption{\it The  probability of correct association $P({\Delta}_{f,c}\geq 0)$ (exact: dashed) and approximated (continuous) versus $N$ ($x$ axis), for various values of $\lambda$: in blue $\lambda=1.5$, in red $\lambda=2.$, in green $\lambda=2.5$. }
\label{ProbaN}
\end{figure}

 Thus, it remains to analyze the effect of the $N$ parameter. This is done in fig.\ref{ProbaN}. Results are restricted to fixed values of $\lambda$, that is equal to $1.5,\,2$ and $2.5$, because they are the most interesting values, representing the more common association problem. We can see that when $N$ exceeds $30$, the approximation is very good. The difference is less than $0.05$, which is quite satisfactory. Moreover, for greater values of $N$, exact values and approximations cannot be distinguished. However, the behavior of the more accurate approximation (see eq. \ref{machine}) is not satisfactory for small values of $N$, since $P({\Delta}_{f,c}\geq 0)(N)$ begins to decrease as $N$ increases.

\begin{figure}[ht!]
\centering
\scalebox{0.35}{\includegraphics{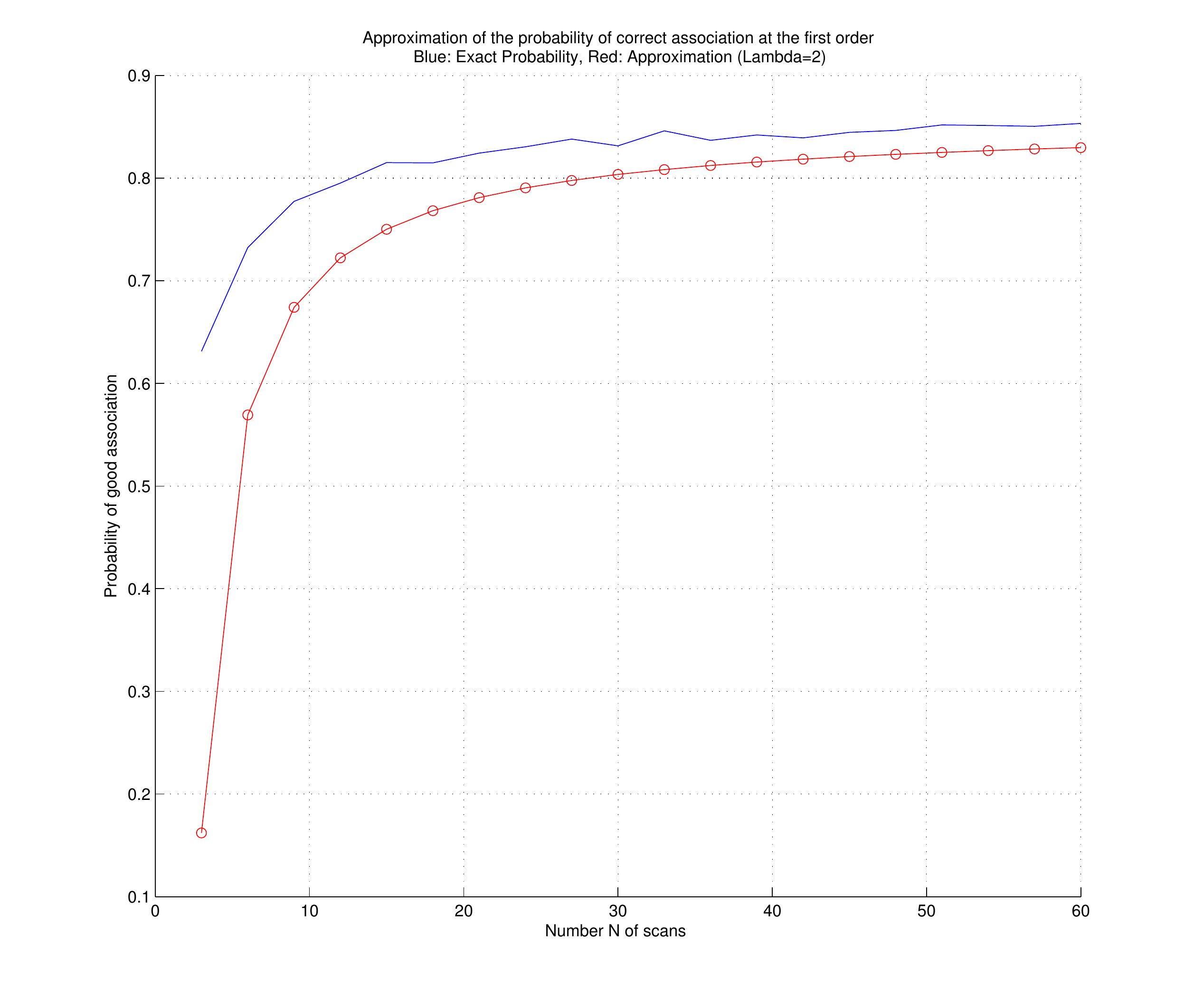}}
\caption{\it The  probability of correct association $P({\Delta}_{f,c}\geq 0)(N)$ versus $N$ ($x$ axis), $\lambda=2$. Blue: exact value, red: $1$-st order approximation (eq. \ref{machine1}).   }
\label{first_order}
\end{figure}

Now, considering the first order approximation of $P({\Delta}_{f,c}\geq 0)(N)$ given by eq. \ref{machine1}, the dependency of $P({\Delta}_{f,c}\geq 0)(N)$ to $N$ is satisfactorily taken into account for "reasonable" values of $N$ (say $10\leq N \leq 40$) , as seen on fig. \ref{first_order}. In particular, the calculated slope (${\sf slo}$, eq. \ref{machine1}) is close to the actual one.

\begin{figure}[ht!]
\centering
\scalebox{0.4}{\includegraphics{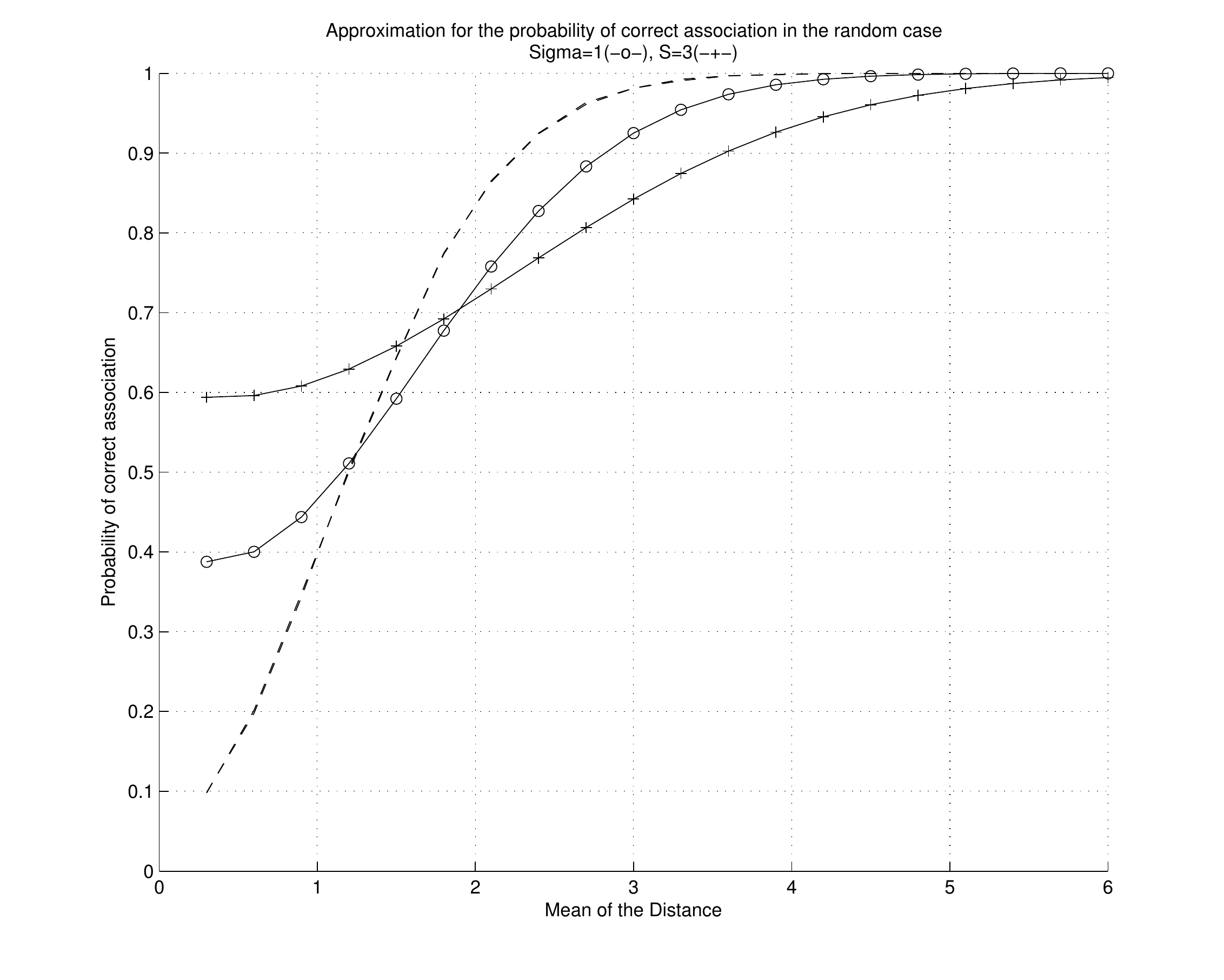}}
\caption{\it The  probability of correct association $P(\bar{\Delta}_{f,c}\geq 0)$ for a random $\lambda$ versus $\bar{\lambda}_{0}$ $x-axis$, $N=40$. Dashed: deterministic $\lambda$ ($\sigma_{0}=0$), continuous: random $\lambda$ ($-o-$: $\sigma_{0}=1$, $-+-$: $\sigma_{0}=3$).}
\label{randomization}
\end{figure}

Finally, we present the results for a random $\lambda$ (see subsection \ref{random_lambda}), on fig. \ref{randomization} . The values of $P(\bar{\Delta}_{f,c}\geq 0)$ are plotted on the $y$-axis, versus the mean value of $\lambda$ ($\lambda_{0}$), for two values of the $\sigma_{0}$ parameters ($1$ and $3$). Not surprisingly,  the effect of this randomization is noteworthy.

\section{The multiple false measurements case}
\label{mfmc}

\begin{figure}[ht!]
\centering
\scalebox{0.30}{\includegraphics{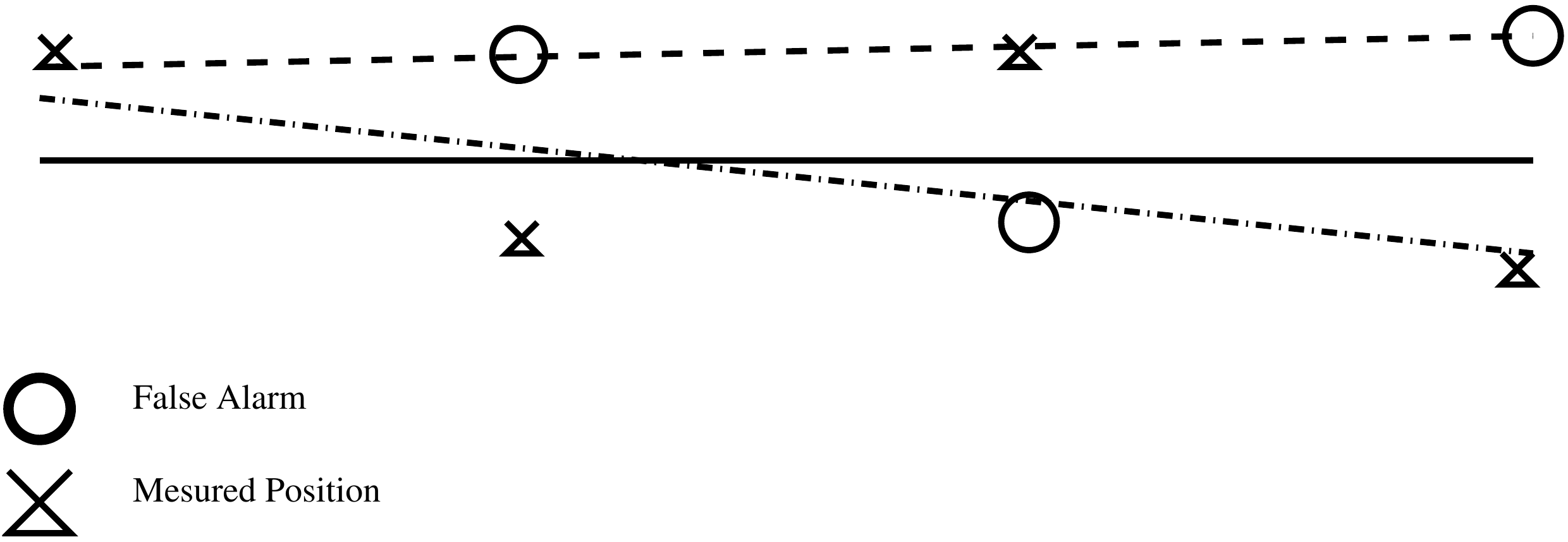}}
\caption{\it The  multiple false measurement scenario }
\label{scenario_multiple2}
\end{figure}

Just like in the first part, a target is moving with a rectilinear and uniform motion. The hypotheses we made in the first part are unchanged. In fact, we consider more specifically the section 3 framework. In this part, we focus on multiple false measurements, and our aim is again to determine the probability for deciding the right association.\\
We have seen previously (see section 3.4) that  a closed form of $\Delta_{f,c}$ could be obtained (see eq. \ref{dpsi}). Thus, calculation of the probability of correct association ($P(\Delta_{f,c} \geq 0$) can be extended to the general case. However, deriving convenient approximations lead us to encounter severe difficulties. So, the feasible approaches will rely on the same principles but with fundamental simplifications. More specifically, we assume that there is at most one false measurement for each time-period.
The  scenario we consider here is depicted on figure \ref{scenario_multiple2}.

In order to investigate the difficulties we have to face, let us consider the numerator of $\Psi_{{\sf{FA}}_{K}}$ (denoted $N(\Psi_{{\sf{FA}}_{K}})$ . Opposite to the unique false measurement case, this numerator cannot be considered (or approximated) by a unique quadratic form (see section 4.2).  Actually, we have (see eq. \ref{dpsi}):
\begin{equation}
N(\Psi_{{\sf{FA}}_{K}})=\displaystyle{\sum_{k=1}^K \sum_{k'=1}^K} \alpha_N(l_{k},l_{k'}) \left( \langle\mathbf{e}_{l_k} , \mathbf{e}_{l_{k'}} \rangle - \langle\sf{fa}_{l_k} , \sf{fa}_{l_k'} \rangle \right)\;.
\end{equation}
A first problem is that $N(\Psi_{{\sf{FA}}_{K}})$ can be small while, simultaneously, elementary terms $\left( \langle\mathbf{e}_{l_k} , \mathbf{e}_{l_{k'}} \rangle - \langle\sf{fa}_{l_k} , \sf{fa}_{l_k'} \rangle \right)$ can be (relatively) large, but of opposite signs. The change of variable approach which is instrumental for deriving explicit closed form approximations of the $B_{i}$ and $A_{i}$ integrals is then clearly unfeasible.\\

So, we have to turn to a {\bf radically} different approach based upon  normal approximations. A key feature of the normal densities is that there are exhaustively represented by their two first moments. Then, we will see that these moments can be easily calculated. In order to give the general scheme, let us recall the general (linear regression) result (see eq. \ref{dpsi}):
\begin{equation}
\begin{array}{l}
\mathcal{L}\left(\Delta_{{\sf{FA}}_{K}}\vert \,\tilde{\beps}_{l_{1}}= \mathbf{e}_{l_{1}} \cdots, \tilde{\beps}_{l_{K}}= \mathbf{e}_{l_{K}}  \right) 
={\mathcal N}\left[  m_{1},v_{1} \right]\;,\\
\mbox{where :}\\
m_{1}=\displaystyle{\sum_{k=1}^K \sum_{k'=1}^K} \alpha_{N}(l_{k},l_{k'}) \left( \langle\mathbf{e}_{l_k} , \mathbf{e}_{l_{k'}} \rangle - \langle\sf{fa}_{l_{k}} , \sf{fa}_{l_{k'}} \rangle \right)\;,\;\\
v_{1}=4 \;\displaystyle{\sum_{k=1}^{K}\sum_{k'=1}^{K} \theta(l_k,l_{k'}) \; \langle\mathbf{e}_{l_k}- {\sf{fa}}_{l_{k}} , \mathbf{e}_{l_{k'}}- {\sf{fa}}_{l_{k'}}}\rangle \;.
\label{normal_m}
\end{array}
\end{equation} 
Assuming that the mean ($m_1$) and the variance ($v_1$) of  $\Delta_{\sf{FA}_{K}}$ are random, thanks to the ($e_{l_k}$) terms, but with determined law, we deduce an expression of the posterior law of the  $\Delta_{\sf{FA}_{K}}$ random variable. More precisely, assume that we have:
$$
 m_1 \sim \mathcal{L}_1(\theta_1) \;\;\mbox{ and:}\;\; v_1 \sim \mathcal{L}_2(\theta_2) \;,
$$
with $\theta_1$ and $\theta_2$ deterministic parameters. Assume also that the density function for $\mathcal{L}_1$ is $g_1$ with support $S_1$ and that for $\mathcal{L}_2$ it is $g_2$ with support $S_2$. Then, the posterior density of  $\Delta_{\sf{FA}_{K}}$ simply reads:
\begin{equation}
\begin{array}{lll}
h\left( \Delta_{\sf{FA}_{K}}\right) &=& \displaystyle{\int_{S_1} \int_{S_2} } f(\Delta \mid m_1,v_1) g_{\theta_{2}}(v_1) g_{\theta_{1}}(m_1) \;d v_1 d m_1 \;.
\end{array}
\label{normal_2}
\end{equation}
The great advantage we have now is that though we do not have the right expression of the posterior law, we just have to consider a double integration. So, the problem we have to face now is  to obtain convenient approximations of $g_{\theta_{1}}$ and $g_{\theta_{2}}$.\\

First, we will approximate the law of the mean $m_{1}$ with a normal distribution. For a great number of random variables,  the central limit theorem allows us to make this approximation. Then, we assume now that $m_1 \sim \mathcal{N}(m_0, \sigma_0^2)$. The distribution of $v_1$ will be discussed {\bf later}. As {\bf both} $\Delta_{\sf{FA}_{K}}$  and $m_1$ are normally distributed, we have a precise knowledge of the posterior density of $\Delta_{\sf{FA}_{K}}$  (see Appendix C):
\begin{equation}
\begin{array}{lll}
\medskip
 h\left( \Delta_{\sf{FA}_{K}}\right) &=& \displaystyle{\int_{S_1} \int_{S_2}} \;f(\Delta_{\sf{FA}_{K}} \mid m_1,v_1) \;g_2(v_1) g_1(m_1) d v_1 d m_1 \;,\\
&=& \displaystyle{\int_{S_2}}\; f_{\mathcal{N}(m_0,\sigma_0^2+v_1)}(\Delta)\; g_2(v_1) d v_1
\end{array}
\label{normal_3}
\end{equation}
Thus, we have:
\begin{equation}
\begin{array}{lll}
P\left(\Delta_{\sf{FA}_{K}} \geq 0 \right) &=& \displaystyle{\int_{S_2} } {\sf{erfc}} \left(\frac{m_0}{\sqrt{\sigma_0^2+v_1}} \right) g_2(v_1) d v_1\;.
\end{array}
\label{normal_4}
\end{equation}
This expression is quite simple and easily computable. Moreover, in this setup, the accuracy of the approximation increases with $K$, thanks to the central-limit theorem. Our problem being to render $h\left( \Delta_{\sf{FA}_{K}}\right) $ (see eq. \ref{normal_3}) as explicit as possible, we have to perform integration w.r.t. the variance $v_1$. To that aim, we have to choose a law for the variance $v_{1}$. We shall consider two solutions:\\

The first one is to use again the central-limit theorem, and to model $v_1$ via a Gaussian distribution\footnote{The limitation of that approach is that if we consider that law, the variance will have non-zero probability to be negative!}. The second solution is to calculate the right law of $v_1$, which should be a kind of Chi-2. \\

Considering the expression of $v_1$, we notice (see eqs. \ref{dpsi}, \ref{normal_m} ) that it is a weighted sum of elementary quadratic forms of normal vectors   ($ \langle \mathbf{e}_{l_{k}}- {\sf{fa}}_{k}, \mathbf{e}_{l_{k'}}- {\sf{fa}}_{k'}\rangle $), with weights $ \theta(l_k,l_{k'})$. Each elementary quadratic form is Chi-square distributed. However, when the weights are different, a tractable distribution of the weighted sum is not available (see \cite{sol}). So, a first simplification is to consider that these weights are approximately equal altogether\footnote{A reasonable assumption, with our assumptions.}. In this setup, we consider that $v_{1}$ is Chi-square distributed with $2K$ degrees of freedom, and we have:
\begin{equation}
\begin{array}{lll}
P\left(\Delta_{\sf{FA}_{K}}\geq 0 \right) &=&\displaystyle{ \int_{\mathbb{R}_{+}}} {\sf{erfc} } \left(\frac{m_0}{\sqrt{\sigma_0^2+v_1}} \right) \;f_{\chi^{2}(2K)}(v_1) d v_1\\\\
{}&=& \frac{1}{2^{K} \Gamma(K)} \displaystyle{\int_{\mathbb{R}_{+}} }{\sf{erfc} } \left(\frac{m_0}{\sqrt{\sigma_0^2+v_1}}\right) \;v_1^{K-1} e^{-v_1/2} d v_1
\end{array}
\label{normal_5}
\end{equation}
Turning now toward the first solution (normal approximation of $v_{1}$, ie $v_1 \sim \mathcal{N}(v_0, s_0^2)$), yields:

\begin{equation}
 P\left(\Delta_{\sf{FA}_{K}}\geq 0 \right)=\displaystyle{\int_{\mathbb{R}_{+}} } {\sf{erfc}}\left(\frac{m_0}{\sqrt{\sigma_0^2+v_1}} \right) \;f_{\mathcal{N}(v_0,s_0^2)}(v_1) d v_1 \;,
\label{normal_6}
\end{equation}
where the parameters $m_0$, $\sigma_0^2$, $v_0$ and $s_0^2$ are given by (see Appendix D):
\begin{equation}
\left|
\begin{array}{lll}
\medskip
m_0 &=& \displaystyle{2\sum_{k=1}^{K} } \alpha_N(l_{k},l_{k}) - \;\displaystyle{\sum_{k=1}^{K} \sum_{k'=1}^{K} } \alpha_{N}(l_{k},l_{k'}) \;\lambda_{l_{k}}\,\lambda_{l_{k'}} \;,\\
\medskip
\sigma_0^2 &=& 4 \; {\left[\displaystyle{\sum_{k=1}^{K} \sum_{k'=1}^{K}} \alpha_{N}(l_{k},l_{k'})\right]}^2\;,\\
\medskip
v_0 &=&  4 \;\displaystyle{\sum_{k=1}^{K} } \theta\left(l_{k},l_{k}\right)\,\left(2 + \lambda^2_{l_k}\right) + 4\; \displaystyle{\sum_{k=1}^{K} }\displaystyle{\sum_{k'=1, k'\ne k}^{K} } \theta(l_{k},l_{k'}) \lambda_{l_k} \lambda_{l_{k'}}\;,\\
\medskip
s_0^2 &=& 2\;\left[\displaystyle{\sum_{k=1}^{K} } \displaystyle{\sum_{k'=1}^{K} } \theta(l_{k},l_{k'})(1+\lambda_{l_k}) (1+\lambda_{l_{k'}})\;\right]\,\left[\displaystyle{\sum_{i=1}^{K} } \displaystyle{\sum_{i'=1}^{K} } \theta(l_{i},l_{i'})\right] \;.
\end{array}
\right.
\label{normal_8}
\end{equation}
 However, even if convenient approximations of the ${\sf{erfc}}(x)$ functions exist, they dont lead, in general, to simple closed form approximations.  So, it seems difficult to obtain a more explicit closed-form approximation for the multiple false measurement case. Some insights can be gained by approximating the $\alpha_N(l_{k},l_{k})$ and $\theta(l_{k},l_{k'})$ (see eqs. \ref{alp-thet}, \ref{dpsi}), under the assumption that the ratio $K/N$ is sufficiently small w.r.t. $1$, yielding:
\begin{equation}
\left|
\begin{array}{ll}
\alpha_N(l_{k},l_{k}) \backsim \left(1-\frac{1}{N}\right)\,K\ &  \theta(l_{k},l_{k'}) \backsim \left(\frac{P(N^3,K^{3})}{N^4} \right) \,K\;,\\
m_0 \backsim  \left(1-\frac{1}{N}\right)\,K^{2}(2-\lambda^{2}\,K)    & \sigma_0^2 \backsim 4 \;K^{6}\;,
\end{array}
\right.
\end{equation}
where $P(N^3,K^{3})$ is a polynomial in $K$ and $N$, whose maximal order in $N$ and $K$ is $3$. Thus, we notice the fundamental importance of the $K$ and $\lambda$ parameters. Similarly to the unique false measurement case (see eq. \ref{machine1}), the effect of $N$ appears as a slope factor toward the steady-state value.

\subsection{Exponential Law Assumption}

We wrote in the previous paragraph:\\
\begin{equation}
\begin{array}{lll}
P\left(\Delta_{\sf{FA}_{K}} \geq 0 \right) &=& \displaystyle{\int_{S_2} } {\sf{erfc}} \left(\frac{m_0}{\sqrt{\sigma_0^2+v_1}} \right) g_2(v_1) d v_1\;.
\end{array}
\end{equation}
We can use the following Taylor development:\\
\begin{eqnarray}
erf(z) &=& \frac{2}{\sqrt{\pi}} \sum_{n=0}^{\infty} \frac{(-1)^n}{n! (2n+1)} z^{2n+1}
\end{eqnarray}
And we then have to calcultate:\\
\begin{equation}
\begin{array}{lll}
P\left(\Delta_{\sf{FA}_{K}} \geq 0 \right) &=&1- \frac{2}{\sqrt{\pi}} \sum_{n=0}^{\infty} \frac{(-1)^n}{n! (2n+1)} \displaystyle{\int_{S_2} }  \left(\frac{m_0}{\sqrt{\sigma_0^2+v_1}} \right)^{2n+1} g_2(v_1) d v_1\;.
\end{array}
\end{equation}
If we assume that $v_1$ follows an exponential law, we then have to calculate that simple integral:\\
\begin{eqnarray}
I_{2n+1} &=& \int_{\mathbb{R}_{+}}\left(\frac{m_0}{\sqrt{\sigma_0^2+v_1}} \right)^{2n+1} v_0 e^{-v_1 v_0} d v_1
\end{eqnarray}
Performing calculations, we then have:\\
\begin{eqnarray}
I_{2n+3} &=& v_0 m_0^{2n+3} - v_0 m_0^{2} \sigma_0^{2n+1} I_{2n+1}
\end{eqnarray}
And then,\\
\begin{eqnarray}
I_{2n+1} &=& v_0^2 m_0^5 \frac{1-(-v_0 m_0^4 \sigma_0 )^n}{1+v_0 \sigma_0 m_0^4} - v_0^n m_0^{2n} \sigma_0^{n-1} I_1
\end{eqnarray}
Which can be used in the sums to calcumate the final expression of the probability:\\
\begin{eqnarray}
P\left(\Delta_{\sf{FA}_{K}} \geq 0 \right) &=& 1- \frac{erf(1)-erf(-v_0 m_0^4 \sigma_0)}{v_0 \sigma_0 m_0^4} \\
&& + \frac{erf(v_0 m_0^2 \sigma_0)}{m_0^2 v_0 \sigma_0^2} I_1
\end{eqnarray}

\section{Simulations: the multiple false measurements case}

\subsection{Multiple false measurements and the probability of correct association}

\begin{figure}[ht!]
\centering
\scalebox{0.60}{\includegraphics{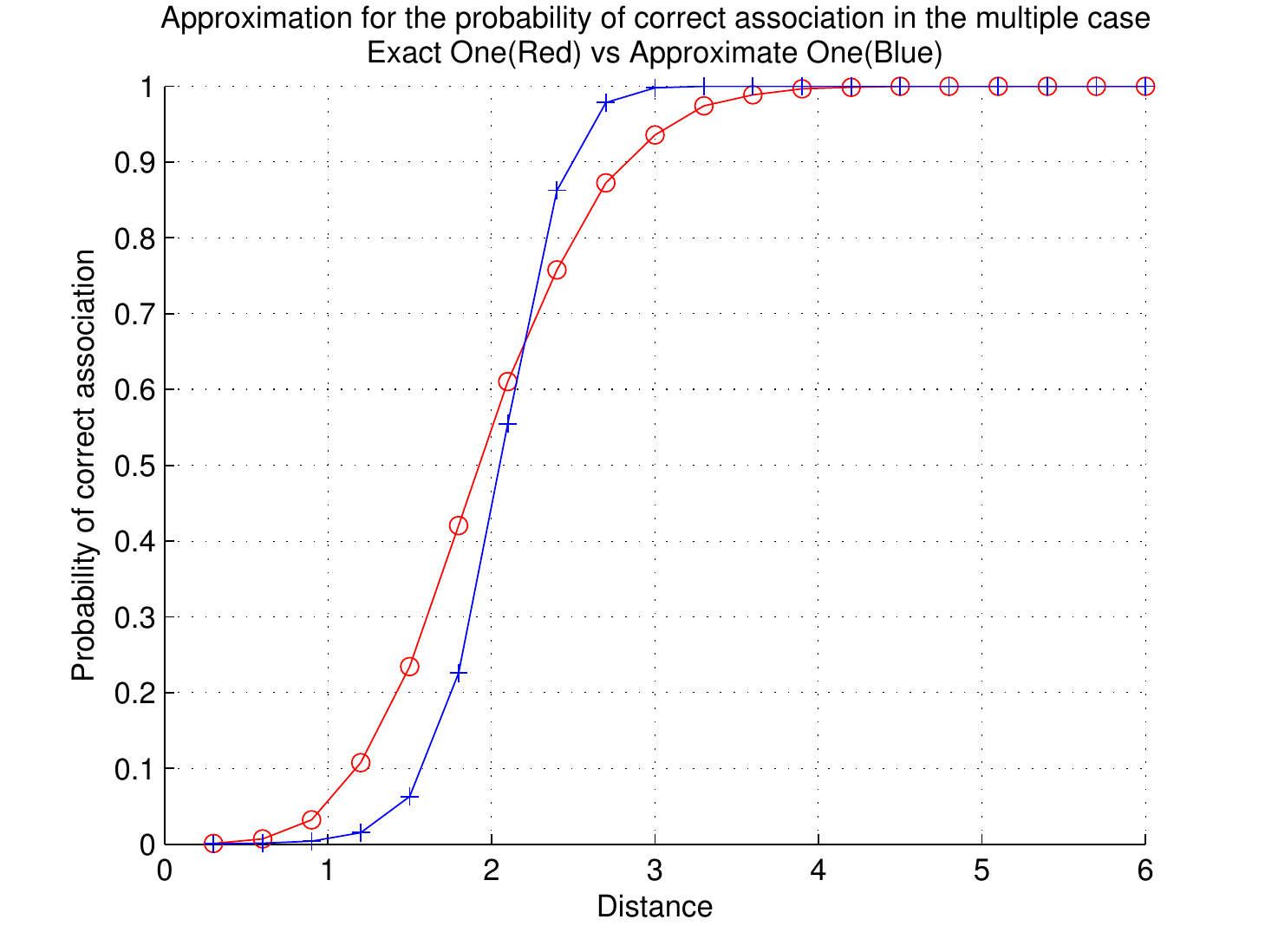}}
\caption{\it Approximation of the Probability of correct association for multiple (consecutive) false measurements ($K=2$, $\chi^{2}$ approximation eq. \ref{normal_5}). $ P\left(\Delta_{\sf{FA}_{K}}\geq 0 \right)$ in the $y$-axis, $\lambda$ on the $x$-axis, $N=40$.}
\label{successive_multiple}
\end{figure}

\begin{figure}[ht!]
\centering
\scalebox{0.60}{\includegraphics{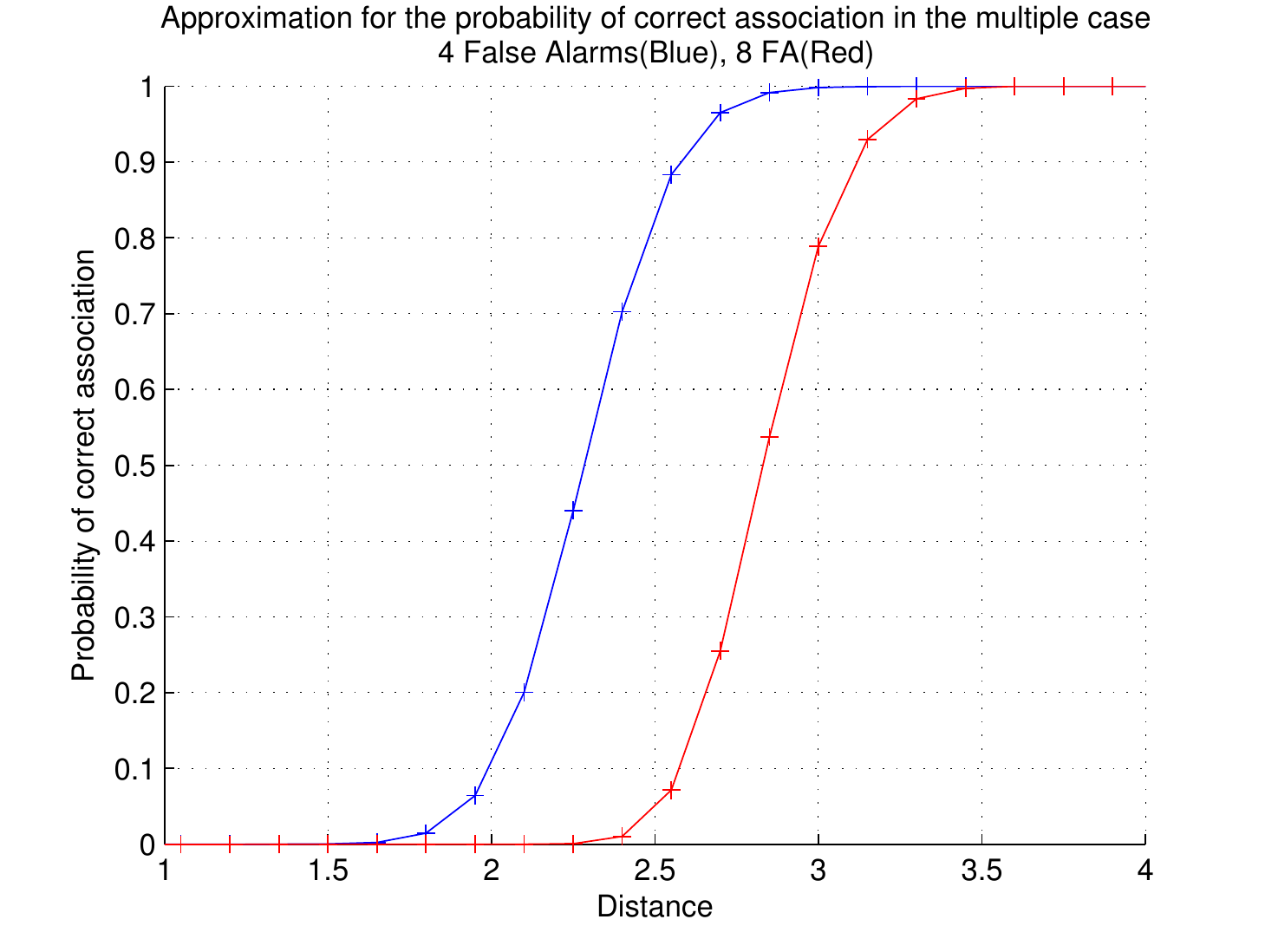}}
\caption{\it Probability of correct association for various number of consecutive false alarms ($K=4$ and $K=8$).  $ P\left(\Delta_{\sf{FA}_{K}}\geq 0 \right)$ in the $y$-axis, $\lambda$ on the $x$-axis, $N=40$.}
\label{successive_multiple_2}
\end{figure}

We consider here the framework which has been develop in the section \ref{mfmc}.
First, we have to consider the validity of the normal ($m_{1}$) and $v_{1}$ approximations (see eqs \ref{normal_5} and \ref{normal_6}). For a value of $K$ ({\small{number of false measurements}}) as small as $2$ and a constant $\lambda$, this is presented in fig. \ref{successive_multiple}, for $N=30$. The result is quite satisfactory, even for this small value of $K$. 
In figure \ref{successive_multiple_2}, we consider the difference between four and eight false measurements. This difference looks like a simple translation. The main result is that having eight false measurements, at a constant distance of $3.5$ is equivalent to a double false measurement scenario, with distance $2.5$ and only one false measurement, with a distance of $1.8$.\\

\section{Conclusion}

Deriving accurate closed-form approximations of the probability of correct association is of fundamental importance for understanding the behavior of data association algorithms. However, though numerous association algorithms are available, performance analysis is rarely considered from an analytical point of view. Actually, this is not too surprising when we consider the difficulties we have to face even in the simplistic framework of linear regression. \\
So, the main contribution of this paper is to show that such derivations are possible. This has been achieved via elementary though rigorous derivations, developed in a common framework. Multiple extensions and applications render it quite attractive for a wide variety of contexts (close targets, clutter, intentionally generated false measurements, ECM, etc.).

\newpage
\section*{Appendix A}
The aim of this appendix is to provide an explicit closed forms of the two quadratic forms defining the mean and the variance of $\mathcal{L}\left(\Delta_{f,c}\vert \,\tilde{\beps}_{l}= \mathbf{e}_{l} \right)$ (see eq. \ref{normal_1}).
The first step consists in calculating a closed form for the $\Psi(\mathbf{e}_l)$ numerator. Considering the special forms\footnote{These two vectors are made of zeros except for $x$ and $y$ $l$-th components} of the vectors $\mathbf{e}_l$ and ${\sf{fa}}_{l}$, only a closed form expression of the $\cam_{l,l}$ ~($2\times2$) $l-$th diagonal block matrix of the $\cam$ matrix is required. Routine calculations yield:
\begin{equation}
\begin{array}{l}
\cam_{l,l} = \left[1-\frac{2\big(2N+1-6\,l+\frac{6\,l^2}{N}\big)}{(N+1)(N+2)}\right]I_2\;,\\
\mbox{so that:}\\
{\mathbf{e}_l}^T \cam \mathbf{e}_l-{\sf{fa}}_{l}^T \cam \;{\sf{fa}}_{l}=\left[1-\frac{2\left(2N+1-6l+\frac{6l^2}{N}\right)}{(N+1)(N+2)}\right]\;\left(\|\mathbf{e}_l\|^2 - \| {\sf{fa}}_{l} \|^2 \right)\;.
\end{array} 
\end{equation} 
In the second step, the $\Psi(\mathbf{e}_l)$ denominator is considered. First, it is worth recalling the form of the $\Phi$ matrix:
\begin{eqnarray} 
\Phi & = & (I-\cah) \Sigma_{{\sf com}} (I-\cah^T) \;,\nonumber\\
& = & \underbrace{\Sigma_{{\sf com}} - \Sigma_{{\sf com}} \cah^T - \cah \Sigma_{{\sf com}}}_{\Phi^{1}} +\cah \Sigma_{{\sf com}} \cah^T\;.
\end{eqnarray} 
Noticing that the $(2\times2)$ sub-matrix  $ \Phi^{1}(l,l) $  is zero, we can restrict to the $(l,l)$ $(2\times2)$ sub-matrix of the $\cah \Sigma_{{\sf com}} \cah^T$ matrix. Straightforward calculations yield:

\begin{equation}
\begin{array}{l}
\cah \Sigma_{{\sf com}} \cah^T=\mathcal{X}\,\mathcal{C} \Sigma_{\sf{com}} \mathcal{C}^T \,\mathcal{X}^{T}\;,\\
\mbox{with:}\\
\mathcal{C}= \left(
\begin{array}{cccc}
(4N+2)I_{2} & \ldots & (4N+2-6(k-1))I_{2}&\ldots \\
-\frac{6}{\delta} I_{2} & \ldots & -\frac{6}{\delta}(1-\frac{2(k-1)}{N})I_{2}&\ldots 
\end{array} 
\right)
\end{array} 
\end{equation} 
For the sake of simplicity, it is assumed that we have  $\Sigma_{\sf{com}}=\mbox{diag}\,\left(\underbrace{I_{2},\cdots,I_{2}}_{l-1}, 0,\underbrace{I_{2},\cdots,I_{2}}_{N-l-1}\right)$. Then, routine calculations yield a simple expression for the $4\times 4$ matrix  $\mathcal{C} \Sigma_{\sf{com}} \mathcal{C}^T$ :

\begin{equation}
\begin{array}{l}
\mathcal{C} \Sigma_{\sf{com}} \mathcal{C}^T=  \frac{1}{(N+1)^2(N+2)^2}
\left(
\begin{array}{cc}
Q_1(l,N)I_2 & Q_2(l,N)I_2\\
Q_2(l,N)I_2 & Q_3(l,N)I_2
\end{array}
\right) \;, 
\end{array}
\end{equation} 
from which, we deduce finally ($\Phi_{l,l}$ $l$-th $2\times2$ diagonal block of the $\Phi$ matrix):

\begin{equation}
\begin{array}{l}
\Phi_{l,l} = \frac{1}{(N+1)^2(N+2)^2} \times \\
\; \left[ Q_1(l,N) + 2l\;\delta Q_2(l,N) + l^2 \; \delta^2 Q_3(l,N) \right]I_{2} \;,
\end{array}
\end{equation} 
where the $Q_1$, $Q_2$ and  $Q_3$ polynomials have the following expression:

\begin{equation}
\left|
\begin{array}{lll}
\medskip
Q_1(l,N) & =& 4N^3-50N^2+N(48l-18)+l(24-36l)+4 \;.\\  \nonumber
\medskip
Q_2(l,N) & = &-\frac{6}{\delta} \left[N^2-5N-2+4l(1+\frac{1}{N}-\frac{3l}{N})\right] \;  \\ \nonumber
\medskip
Q_3(l,N) & = & \frac{36}{\delta^2} \left[ \frac{N}{3}-1+\frac{2}{N}(\frac{1}{3}+2\,l-\frac{2\,l}{N^2}) \right] \;.
\end{array}
\right. 
\end{equation} 

Finally, we have thus obtained:
\begin{eqnarray}
\hspace{-1cm}(\mathbf{e}_{l}-{\sf{fa}}_{l})^{T} \Phi (\mathbf{e}_{l}-\sf{fa}_{l}) &=& \frac{{\|\mathbf{e}_l-{\sf{fa}}_{l}\|}^{2}}{(N+1)^2(N+2)^2} \times \\ 
&&\left[ Q_1(l,N) + 2l\;\delta Q_2(l,N) + l^2 \; \delta^2 Q_3(l,N) \right] \nonumber 
\end{eqnarray}

\section*{Appendix B}
This appendix deals with the calculation of the coefficients $\gamma_{i}$ for the least square criterion. Denoting $\varphi_{i}$  ($i=1,\cdots,n$) the functions defined by $\varphi_{i}\stackrel{\Delta}{=}\frac{n}{6 i \;\sf{den}} \,\mathbf{1}_{[b_{\sf{inf}}^{i},b_{\sf{sup}}^{i}]}$, the coefficients $\gamma_{i}$ are the solutions of the following optimization problem:

\begin{equation}
\displaystyle{\min_{\gamma_{i}} }\;{\|g-\sum_{i=1}^{n}\gamma_{i} \varphi_{i} \|^{2}_{2}}\;,
\label{opt_1}
\end{equation}
where $g$ is the normal density given by eq. \ref{normal_m}, and $\|-\|_{2}$ is the $L^{2}$ norm. It is then known that the $\gamma_{i}$ are the solutions of the following linear system:
{\small
\begin{equation}
\left\{
\begin{array}{lll}
\gamma_{1} {\|\varphi_{1}\|}^{2}_{2}+\gamma_{2}\langle \varphi_{2},\varphi_{1}\rangle+\cdots+\gamma_{n}\langle \varphi_{n},\varphi_{1}\rangle&=&\langle g, \varphi_{1} \rangle \;,\\
\vdots & &\\
\gamma_{1}\langle \varphi_{1},\varphi_{n}\rangle+\gamma_{2}\langle \varphi_{2},\varphi_{n}\rangle+\cdots+\gamma_{n} {\|\varphi_{n}\|}^{2}_{2}&=&\langle g, \varphi_{n} \rangle \;.
\end{array}
\label{sys-g}
\right.
\end{equation} }
The norms ${\|\varphi_{i}\|}^{2}_{2}$, as well as the scalar products $\langle \varphi_{i},\varphi_{j}\rangle$ are straightforwardly calculated , yielding:
\begin{equation}
\langle \varphi_{i},\varphi_{j}\rangle=\frac{n}{6\;\inf(i,j)}\;\frac{1}{\sf{den}}\;.
\end{equation}
and solving the linear system given by eq. \ref{sys-g}:
\begin{equation}
\displaystyle{\sum_{i}^{n} }\gamma_{i}=\langle g, \mathbf{1}_{[b_{\sf{inf}}^{1},b_{\sf{sup}}^{1}]}\rangle\;\\
\gamma_{i}=i(i-1) \langle g, \varphi_{i-1}-\varphi_{i} \rangle -i(i+1) \langle g, \varphi_{i}-\varphi_{i+1} \rangle\;.
\label{coeff-g}
\end{equation}
Then, from the above equation (eq. \ref{coeff-g}), we have:
\begin{equation}
\left\{
\begin{array}{l}
\displaystyle{\sum_{i=1}^{n}} \;i\gamma_{i}=2\;\langle g, \displaystyle{\sum_{i=1}^{n}} i\;\mathbf{1}_{[b_{\sf{inf}}^{i},b_{\sf{sup}}^{i}]}\rangle\;,\\
\displaystyle{\sum_{i=1}^{n}} \;\frac{\gamma_{i} }{i}= \frac{1}{n} \;\langle g, \mathbf{1}_{[b_{\sf{inf}}^{n},b_{\sf{sup}}^{n}]}\rangle \;.
\end{array}
\right.
\label{coeff-sum}
\end{equation}
From eq. \ref{coeff-sum}, we deduce the slope of $P(\bar{\Delta}_{f,c}\geq 0)$ as a function of $N$ (see eq. \ref{machine1}):
\begin{eqnarray}
{\sf{slo}} & = & \frac{6}{n} \;\displaystyle{\sum_{i=1}^{n}} \;i\gamma_{i}-\displaystyle{\sum_{i=1}^{n}} \;\frac{\gamma_{i} }{i}\;,\\
&=& \frac{1}{n} \left(12\;\langle g, \displaystyle{\sum_{i=1}^{n}} i\;\mathbf{1}_{[b_{\sf{inf}}^{i},b_{\sf{sup}}^{i}]}\rangle-\langle g, \mathbf{1}_{[b_{\sf{inf}}^{n},b_{\sf{sup}}^{n}]}\rangle \right)\;.
\label{slo-g}
\end{eqnarray}
Obviously, the slope $\sf{slo}$ is positive (see eq. \ref{slo-g}).

\section*{Appendix C}
Here, our iam is simply to recall a classical statistical result.  Assume that the random variable $X$ has the following (conditional) distribution:
\begin{equation}
X \mid m \sim \mathcal{N}(m,\sigma^2)\;,
\end{equation}
with $ m \sim \mathcal{N}(\theta, s^2)$.  Then, integrating over $m$, we have:
\begin{equation}
\begin{array}{lll}
\medskip
h(x) &=&\displaystyle{ \int_{\mathbb{R}}} f(x \mid m) g(m) \;d m  \;, \\ 
{} &=& \displaystyle{\int_{\mathbb{R}} } \frac{1}{2 \pi \sigma s} e^{- \big( \frac{x-m}{\sqrt{2 \sigma^2}} \big)^2 - \big( \frac{m-\theta}{\sqrt{2 s^2}} \big)^2} dm\\
\end{array}
\label{appA_1}
\end{equation}
Performing the integration w.r.t. the $m$ parameter is quite easy since it involves a quadratic form in $m$ and the result is as simple as:

\begin{equation}
h(x)=\frac{1}{2 \pi (s^{2}+\sigma^{2})} \exp\left[-\frac{1}{2(s^{2}+\sigma^{2})} {\left(x-\theta\right)}^{2} \right]\;,
\end{equation}
which shows that the random variable $X$ is normally distributed, with mean $\theta$, and variance $(\sigma^2+s^2)$. So, the uncertainty in the mean $m$ simply results in an incresed variance.

\section*{Appendix D}
The aim of this appendix is the calculation of the values of $m_0$, $\sigma_0^2$, $v_0$ and $s_0^2$. Calculations are a bit long but elementary, so we then just express here the main stages to perform the results. First, we have:

\begin{equation}
\left|
\begin{array}{lll}
\medskip
m_0 &=& \mathbb{E}_{XY} \left[ \displaystyle{\sum_{k=1}^{K} }\displaystyle{\sum_{k'=1}^{K} } \alpha_{N}(l_{k},l_{k'}) (x_{l_{k}} x_{l_k'}+y_{l_k}\,y_{l_k'}-\lambda_{l_k}\,\lambda_{l_k'}) \right] \;,\\
\medskip
\sigma_0^2 &=& \mathbb{V}_{XY} \left[  \displaystyle{\sum_{k=1}^{K} }\displaystyle{\sum_{k'=1}^{K} } \alpha_{N}(l_{k},l_{k'}) (x_{l_k}\,x_{l_k'}+y_{l_k}\,y_{l_k'}-\lambda_{l_k}\,\lambda_{l_k'}) \right] \;,\\
\medskip
v_0 &=& 4\mathbb{E}_{XY} \left[\displaystyle{\sum_{k=1}^{K} }\displaystyle{\sum_{k'=1}^{K} } \theta(l_{k},l_{k'}) \left((y_{l_k}-\lambda_{l_k})\,(y_{l_k'}-\lambda_{l_k'})+x_{l_k}\,x_{l_k'} \right) \right] \;,\\
\medskip
s_0^2 &=& 16\mathbb{V}_{XY} \left[\displaystyle{\sum_{k=1}^{K} }\displaystyle{\sum_{k'=1}^{K} } \theta(l_{k},l_{k'}) \left( (y_{l_k}-\lambda_{l_k})(y_{l_{k'}}-\lambda_{l_k'})+x_k\,x_{k'} \right) \;\right]\;.
\end{array}
\right.
\label{normal_7}
\end{equation} 

These calculations are routine exercises, only the last calculation require (a bit) more attention. In the independent case:
\begin{equation}
\begin{array}{lll}
\mathbb{V}_{XY}(xy) &=& \mathbb{V}(x) \mathbb{E}(y^2) + \mathbb{V}(y) \mathbb{E}(x^2)\;.
\end{array}
\label{appB_4}
\end{equation}
The (small) problem we have to solve is the calculation of the second term. This is achieved via classical results about moments of a normal random variable: 
\begin{equation}
\begin{array}{lll}
\mathbb{V}_{Y} \big[(y_k-\lambda_k)^2 \big] &=& \mathbb{E}\big[ (y_k-\lambda_k)^4 \big] - \mathbb{E}^2 \big[ (y_k-\lambda_k)^2 \big]\;,\\
{} &=& \mathbb{E}\big[ y_k^4-4\,y_k^3 \lambda_k + 6\,y_k^2 \lambda_k^2 - 4\,y_k \lambda_k^3 + \lambda_k^4 \big] - (1+\lambda_k^2)^2 \;,\\
{} &=& 3+6\lambda_k^2+\lambda_k^4-1-2\lambda_k^2-\lambda_k^4 \;,\\
{} &=& 2+4\lambda_k^2 \;.
\end{array}
\label{appB_5}
\end{equation}
Finally, we have:
\begin{equation}
\begin{array}{lll}
s_{01}^2 &=& 64 \sum_{k=1}^{K} \theta^2(k,k) (1+\lambda_k^2)
\end{array}
\label{appB_6}
\end{equation}

\begin{IEEEbiography}[{\includegraphics[scale=0.5]{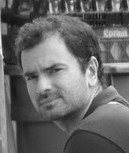}}]{Adrien Ickowicz}
received his PhD degree in Statistics in 2010, under the supervision of J.-P. Le Cadre and then F. Le Gland, from Universit\'{e} de Rennes 1. He is currently a post-doctoral fellow in CMIS, CSIRO in North Ryde, Australia. Before joining CSIRO he lectured at Universit\'{e} de Paris-Dauphine and worked as a CNRS research fellow at Universite de Lille. His research interests includes space-state models, applied time series, computational statistics and Bayesian statistics along with the related applications in different areas such as target tracking, survival analysis, signal processing.
\end{IEEEbiography}
\begin{IEEEbiographynophoto}{Jean-Pierre Le Cadre}
 received the M.S. degree in Mathematics in 1977, the "Doctorat de 3-eme cycle" in 1982 and the "Doctorat d'Etat" in 1987, both from INPG, Grenoble, France.
From 1980 to 1989, he worked at the GERDSM (Groupe d'Etudes et de Recherche en Détection Sous-Marines), a laboratory of the DCN (Direction des Constructions Navales), mainly on array processing. In this area, he conduced both theoretical and practical researches. In particular, he participated to the pratical evaluation of high resolution methods on real data (towed arrays).
Since 1989, he had been with IRISA / CNRS, where he was a CNRS (National Center for Scientific Research) "Directeur de recherche". At that time, his interests moved towards other topics like system analysis, detection, data association and operations research.  
He was awarded Automatica Outstanding reviewer in 2005 and received two prestigious prizes: the Eurasip Signal Processing Best Paper Award (1993) and the IEEE Barry Carlton Award (2008). He was also member of various societies of IEEE.
\end{IEEEbiographynophoto}
\end{document}